\documentclass[superscriptaddress,groupedaddress,nofootnoteinbib,11pt]{article}
\newcommand{\mpl}{M_{\mathrm{Pl}}}

\usepackage{jheppub} 

\usepackage{amsmath, amsfonts, amsthm, amssymb, graphicx, color, hyperref}

\usepackage{float, subfig}
\usepackage{pstool}
\usepackage[utf8]{inputenc}
\usepackage[normalem]{ulem}
\usepackage[english]{babel}
\usepackage{blkarray}
\usepackage{dsserif}
\usepackage{mathtools}
\newcommand{\stu}{St\"uckelberg }
\usepackage{makecell}
\usepackage{multicol}
\usepackage{cleveref}
\usepackage{url}
\usepackage{hyperref}

\def\d{\mathrm{d}}

\def\({\left(}
\def\){\right)}

\def\mpl{M_{\rm Pl}}
\def\p{\partial}

\def\X{\mathcal{X}}

\def\nn{\nonumber}

\def\d{\mathrm{d}}

\def\({\left(}
\def\){\right)}
\def\nn{\nonumber}
\def\p{\partial}

\newcommand{\ie}{{\it i.e.,}\ }

\allowdisplaybreaks[3]

\def \bal#1\eal  {\begin{align} #1 \end{align}}
\def\({\left(}
\def\){\right)}
\def\[{\left[}
\def\]{\right]}
\def\<{\langle}
\def\>{\rangle}
\def\d{\mathrm{d}}

\def\nn {\nonumber}

\def\be{\begin{equation}}
	\def\ee{\end{equation}}
\def\ba{\begin{eqnarray}}
	\def\ea{\end{eqnarray}}
\def\d{\mathrm{d}}

\usepackage[dvipsnames]{xcolor}
\definecolor{c1}{HTML}{D9A1A7}
\definecolor{c2}{HTML}{A1B4A5}
\definecolor{c3}{HTML}{FCD38F}
\definecolor{c4}{HTML}{309F96}

\title{Diffeomorphic Scalar Duality}

\author[a]{Claudia de Rham,}
\author[a]{and Andrew J. Tolley}

\affiliation[a]{Abdus Salam Centre for Theoretical Physics, Imperial College, London, SW7 2AZ, UK}

\emailAdd{c.de-rham@imperial.ac.uk}
\emailAdd{a.tolley@imperial.ac.uk}

\preprint{{\footnotesize Imperial--TP--2026--cdr--2~~~~~~~~ }}

\date{\today}

\abstract{  
We show that every local scalar effective field theory 
admits a new kind of duality to an infinite class of local scalar field theories with distinct Lagrangians. The duality map takes the form of a field-dependent diffeomorphism, and cannot be obtained via purely local field redefinitions, nevertheless the dual theory has an identical $S$-matrix. The subset of interactions that maintain second-order equations of motion is non-trivially mapped into themselves under this transformation. 
We show how to couple generic scalar field theories to gravity in a way that preserves the duality. Crucially, this requires working in the Einstein-Cartan formalism, with the vielbein and spin connection treated as independent variables. When coupling to massless gravity, the duality is interpreted as a local field redefinition in which the vielbein transforms while the spin connection is held fixed; consequently, a torsion-free configuration is generically mapped to a dual configuration with non-zero torsion. 
We specify the general family of first-order gravitational theories that map into themselves under the duality. In the weak gravitational field limit, these reduce to scalar theories kinetically mixed with the graviton, which themselves form a family closed under the duality.
}

\DeclareUnicodeCharacter{202F}{\,}
\begin{document}

\maketitle

\section{Introduction}

Scalar field theories are among the most thoroughly explored and well understood theoretical tools given their relative simplicity and their broad applicability in contexts such as the Higgs mechanism, cosmological inflation, Ginzburg–Landau theory, solitons, dilatons, higher-dimensional moduli, etc\footnote{The work presented here is also trivially generalisable to axions and other pseudo-scalar fields, but we shall focus on scalar fields for concreteness purposes}. Given this, it is tempting to assume that they contain few remaining surprises. In particular, for a real scalar with no global or local symmetries, it would seem that there are few interesting phenomena beyond that which can be inferred by inspection of the form of its Hamiltonian. Given the apparent triviality of $\lambda \phi^4$ in $d=4$, the modern understanding of scalar theories is that they are generically effective field theories, valid over a range of energy scales, and accompanied by an infinite tower of higher derivative interactions suppressed by the cutoff of the EFT. Thus, generic scalar EFTs will admit derivative interactions. A well-known class of such models with a wide range of applications is those for which the Lagrangian is well approximated as a generic function of the scalar and the canonical scalar kinetic term. These arise as low energy descriptions of D-branes \cite{Leigh:1989jq}, as in the DBI action, or as effective descriptions of fluids and superfluids \cite{Son:2002zn,Endlich:2010hf,Dubovsky:2011sj}. \\

In a situation where a local symmetry is broken, it can always be restored using a set of \stu fields. In the global limit, the \stu fields are the Goldstone modes for the broken global symmetry.
When the local symmetry in question is diffeomorphism invariance, the \stu fields are a set of up to $d$ scalar fields $\Phi_a$ which can be regarded in the global limit as the Goldstone modes associated with broken spacetime translations. The global index $a$ can be naturally associated with a locally inertial co-ordinate basis $x^{\mu}$, \ie $a \leftrightarrow \mu$, for which we can identify the longitudinal part of the scalar \stu fields as $\Phi^L_a=\p_a \phi$. If the underlying theory has a global Poincar\'e symmetry, then the following situation naturally occurs:

$
\begin{aligned} \notag
\hspace{-0.7cm}\boxed{
\begin{minipage}{0.45\textwidth}
\begin{center}
 $\Phi_\mu^L=\p_\mu \phi$ transforms as a \textbf{Vector} \\ 
 under global Poincar\'e,
 \end{center}
\end{minipage}
\text{and}
\begin{minipage}{0.48\textwidth}
\begin{center}
  $\Phi_\mu^L=\p_\mu \phi$ transforms as a \textbf{Scalar} \\ 
 under $L$-diffs $\p_\mu \phi \to  \p_\mu \phi$
  \end{center}
\end{minipage}
}
\end{aligned}\vspace{0.4cm}
$

\noindent By $L$-diffs, we mean a subgroup of local diffeomorphisms that preserve the longitudinal nature of $\Phi^L_{\mu}$\footnote{Clearly a generic diffeomorphism of $\Phi^a_L$ will generate transverse vector components. However, as we shall see, a subgroup can be defined which do not.}. Since the derivative of $\phi$ transforms as a scalar, $\phi$ itself is not a diffeomorphism scalar in the usual sense. The apparent mismatch between the global and local transformation properties is a direct result of the symmetry breaking.

In this work, we show that every scalar EFT in Minkowski spacetime—irrespective of its origin, physical interpretation, or the presence of global symmetries—possesses an infinite number of dual formulations with the following transformation properties: \\

$
\begin{aligned} \notag
\hspace{-0.5cm}\boxed{
\begin{minipage}{0.35\textwidth}
\begin{center}
 $\p_\mu \phi$ transforms as a \textbf{Vector} \\ 
 under global Poincar\'e,
 \end{center}
\end{minipage}
\qquad \text{and} \qquad
\begin{minipage}{0.45\textwidth}
\begin{center}
$\p_\mu \phi$ transforms as a {\bf Conformal Scalar}
  $\p_\mu \phi \to \Omega(\phi, (\p \phi)^2) \p_\mu \phi$  
  \end{center}
\end{minipage}
}
\end{aligned}\vspace{0.4cm}
$

\noindent  The precise transformation can be viewed as a combination of a local field redefinition coupled with a field-dependent diffeomorphism of the spacetime coordinates. Given this we refer to it as a {\it Diffeomorphic Scalar Duality}. 
A summary of the duality transformations is given in Table~\ref{tab:dualitysummary}.
This duality differs from those that arise in supergravity and string theory \cite{Hull:2025jpv}, although a better understanding of their potential connection with  superstring field theory \cite{Hull:2025mtb} would be interesting. The Diffeomorphic Scalar Duality we propose here is a true duality in this sense that the dual theory is naively distinct, and cannot be obtained by a local field redefinition from the original, nevertheless it gives an equivalent description of physics, and in particular leads to an identical S-matrix\footnote{Invariance at tree-level is `trivially' guaranteed and we discuss invariance at loop-level in Section~\ref{Smatrix}.}.

The duality falls out of the regime of applicability normally considered by the equivalence theorem. Explicitly, the duality transformation is non-local. A special case of this duality was first noticed for the class of Galileon theories \cite{deRham:2013hsa,deRham:2014lqa,Curtright:2012gx} and DBI-Galileon theories \cite{Chagoya:2016jyn}, both of which are scalar theories with nonlinearly realised spacetime symmetries. In the above language, those examples correspond to $\Omega=1$, for which $\p_\mu \phi$ transforms as a scalar under the new transformation, even though $\phi$ is a scalar under global Poincar\'e.
Our generalisation to all scalar theories, further allowing for a generic conformal factor $\Omega$ (subject to an integrability condition), shows that this nonlinearly realised symmetry plays no role in the existence of the duality, and is largely incidental. The only interesting feature of the nonlinearly realised symmetry is that there is a subclass of duality transformations that preserve manifest symmetry at the level of the action.

\begin{figure}
    \centering
    \includegraphics[width=0.75\linewidth]{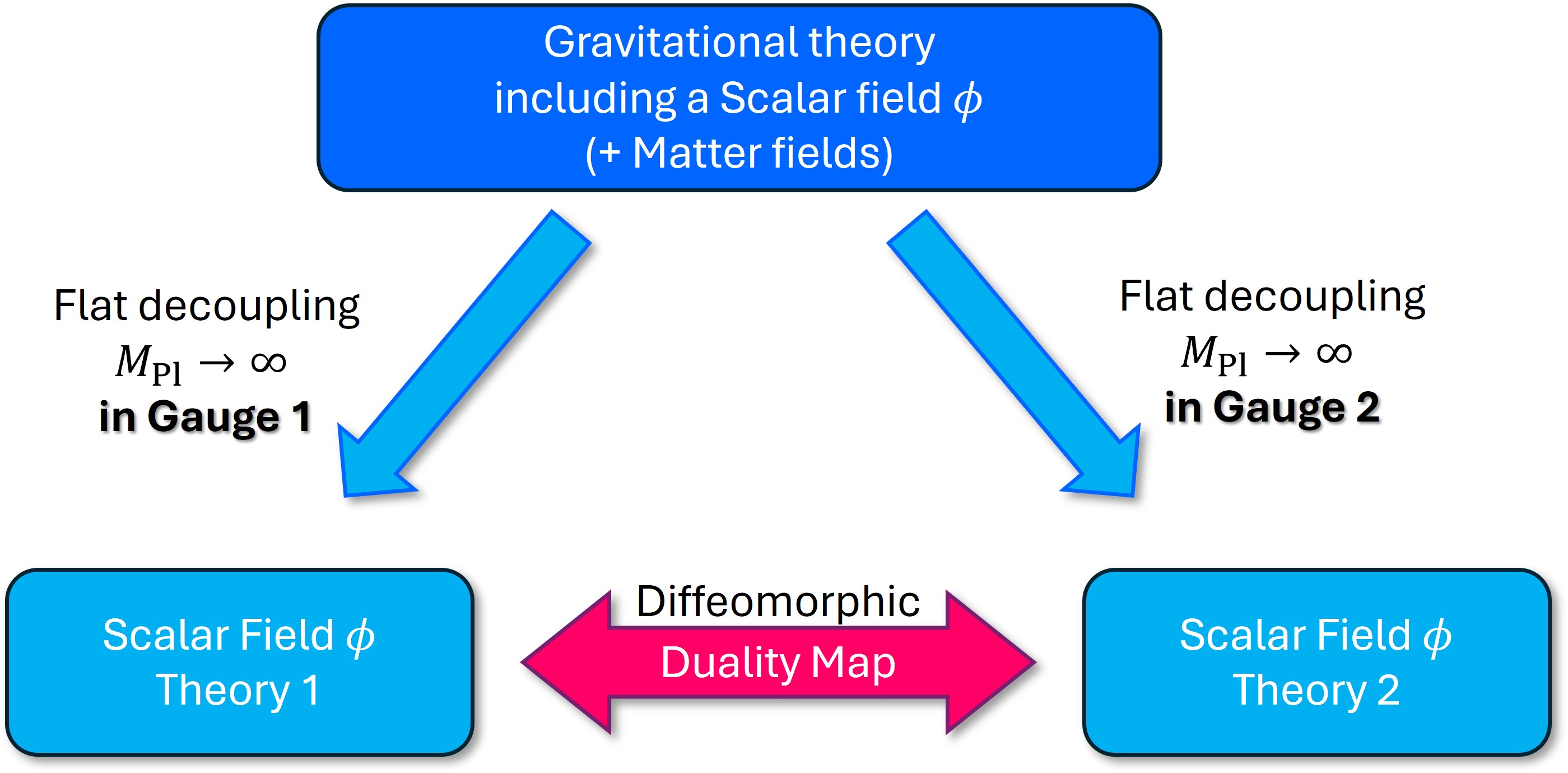}
    \caption{The Diffeomorphic Scalar Duality can be viewed as arising from two different decoupling limits, associated with two different gauge choices, of the same underlying gravitational theory. In the decoupling limit the background geometry becomes Minkowski, but the two Minkowski geometries are related by a field-dependent diffeomorphism.}
    \label{fig:placeholder}
\end{figure}

We shall show how to couple the scalars to massless Einstein gravity in a way that preserves the duality\footnote{For related work using the coset construction in the more restrictive case of Galileon duality see \cite{Baratella:2015yya}.}. A summary of the duality transformations in the gravitational context is also given in Table~\ref{tab:dualitysummary}. Since massless gravity is diffeomorphism invariant, the field-dependent diffeomorphism becomes unnecessary. Rather, the duality is interpreted as a purely local field redefinition of the metric/vielbein and scalar field and any other matter. Given the local gravitational theory, we can always take a decoupling limit via $\mpl \rightarrow \infty$ ($G \rightarrow 0$) which linearises the metric fluctuations. In taking this decoupling limit, it is necessary to specify a coordinate system for the emergent background Minkowski spacetime. Making different choices of coordinates, or equivalently different choices of vielbeins, leads to superficially distinct decoupling limit theories which describe a scalar coupled via mixing terms to a massless graviton. The duality symmetry between the naively distinct decoupling limits emerges as a remnant of the diffeomorphism invariance of the underlying gravitational theory. This is in exact analogy with how the duality was first observed to arise in the context of massive bigravity theories \cite{BiHiguchi}. However, unlike in the massive gravity context, no spontaneous breaking of the diffeomorphism symmetry is required to see the emergence of the duality. Figure \ref{fig:placeholder} illustrates this basic connection. 

Within the gravitational theory, the duality amounts to a purely local, invertible field redefinition that maps single particle asymptotic states into themselves. As such, it satisfies the requirements of the equivalence theorem (equivalence under field redefinitions) \cite{Kamefuchi:1961,Chisholm:1961,Salam:1970,Borchers1960}. This implies that all tree-level scattering amplitudes are left invariant by the duality transformation (invariance at loop-level is discussed in Section~\ref{Smatrix}). In the decoupling limit $\mpl \rightarrow \infty$, the graviton that contributes to scalar exchanges drops out, and we recover the scattering amplitudes of the pure scalar theory. These inevitably inherit invariance under duality at the classical level, and it is straightforward to demonstrate this in examples \cite{deRham:2013hsa}. The reason this holds is that although the duality transformation is non-local acting on the scalar fields, it is perturbatively local and so the infinitesimal form of the duality transformation is local and does satisfy the requirements of the equivalence theorem. 

We begin in Section~\ref{Galileons} with a review of a class of theories which are already known to exhibit the duality, and in Section~\ref{GalileonDuality} review the Galileon duality in a manner which will be easily generalisable. The generalisation is then explicitly spelt out in Section~\ref{GenericScalar}. In Section~\ref{Mattercoupling} we show how to couple the scalars to generic matter fields of arbitrary spin in a way that preserves duality. Then in Section~\ref{gravity} we explain how to extend the duality to gravitational theories and, in turn, how the gravitational theories provide a natural explanation of the existence of the duality via decoupling limits, as summarised by Figure~\ref{fig:placeholder}. Finally, in Section~\ref{Smatrix} we discuss the invariance of the S-matrix and path integral measure.

 \section{Nonlinearly Realised Spacetime Symmetries}

\label{Galileons}

\subsection{Galileons}

 A particularly simple example to start with is the family of scalar theories that nonlinearly realise additional spacetime symmetries, the most well-known examples being Galileons \cite{Nicolis:2008in} and Dirac-Born-Infeld probe-brane scalar theories \cite{Leigh:1989jq}. A Galileon is a local field $\pi(x)$ that is invariant under the nonlinearly realised transformation 
 \ba
 \label{eq:GalShift}
 \pi(x) \rightarrow \pi(x) + c + v_{\mu} x^{\mu}\,.
 \ea
Such fields originally emerged in the context of the Dvali-Gabadadze-Porrati model \cite{Dvali:2000hr,Dvali:2000rv} by considering a particular decoupling limit \cite{Luty:2003vm}, and it was subsequently realised that Galileons inevitably arise in the decoupling limit of massive gravity theories \cite{deRham:2010ik,deRham:2007xp,deRham:2010gu,deRham:2011ca,deRham:2010kj,deRham:2010tw, deRham:2011qq,Ondo:2013wka,BiHiguchi,deRham:2023byw}, where the Galileon field $\pi(x)$ has the clear interpretation as the helicity-zero mode of the massive graviton mode.

The Galileon symmetry can also be viewed as arising from a Wigner--\.{I}n\"{o}n\"{u} contraction of the Poincar\'e symmetry of higher dimensional brane effective theories \cite{deRham:2010eu} (see \cite{Hinterbichler:2010xn,Trodden:2011xh,Garoffolo:2025igz} for its multi-field generalisation), or it can be viewed on its own terms as a nonlinearly realised symmetry via the coset construction \cite{Goon:2012dy}. In its massive gravity construction, the symmetry is associated with translations in the global Poincar\'e group of the reference Minkowski metric. There has been considerable work on extensions, such as the de Sitter Galileon \cite{Burrage:2011bt} which incorporates an extended version of the Galileon symmetry, or consideration of coupling to gravity \cite{Deffayet:2009wt,Gabadadze:2012tr,Andrews:2013ora,Goon:2014ywa}.

As an EFT, the Galileon exhibits several interesting properties: There exists a non-renormalisation theorem for its leading interactions \cite{Luty:2003vm,Nicolis:2004qq,deRham:2012ew,deRham:2013qqa,Goon:2016ihr}, and it naturally exhibits the Vainshtein mechanism inherited from massive gravity  \cite{Vainshtein:1972sx,Babichev:2013usa,deRham:2014zqa,deRham:2023byw}. A special class of Galileons \cite{Hinterbichler:2015pqa} are known to play an important role in scattering amplitudes \cite{Cachazo:2014xea,Cheung:2014dqa,Carrasco:2019qwr} and they provide interesting examples of technically natural inflation \cite{Burrage:2010cu}. On the one hand, Galileon theories are known to violate the usual S-matrix positivity bounds \cite{Tolley:2020gtv}, indicating that if they possess a UV completion, it must be non-standard. On the other hand, the kinds of interactions characteristic of Galileons do appear within ordinary effective field theories that do satisfy S-matrix positivity constraints.
In our present context, we are interested in the Galileon only as a worked-out example of the diffeomorphic scalar duality, and the duality transformation we consider applies equally well to theories which do satisfy S-matrix positivity requirements.
 
Among Galileon field theories that are invariant under the nonlinearly realised transformation \eqref{eq:GalShift}, there exists a privileged class of operators which lead to second order equations of motion which, in the EFT context, will dominate the long wavelength dynamics for fluctuations around any background. These may be written in a number of different forms and it shall prove useful to consider each of them. Arguably, the most elegant form is
 \be \label{gal1}
S_1 = \int \d^d x \[ \sum_{n=0}^d c_n \pi \, \epsilon \epsilon \eta^{d-n} (\partial \partial \pi )^{n} \] \, .
 \ee
We have used here the shorthand notation for two 2-tensors $A_{\mu\nu}$, $B_{\mu\nu}$
\be
\epsilon \epsilon A^{d-n} B^n \coloneqq \epsilon^{\mu_1 \dots \mu_d}\epsilon^{\nu_1 \dots \nu_d} A_{\mu_1 \nu_1} \dots A_{\mu_{d-n} \nu_{d-n}}B_{\mu_{d-n+1} \nu_{d-n+1}} \dots B_{\mu_{d} \nu_{d}} \, ,
\ee
with $\epsilon$ the Levi-Civita symbols and in the present case $A_{\mu\nu}=\eta_{\mu\nu}$, $B_{\mu\nu}=\partial_{\mu} \partial_{\nu} \pi$. The virtue of the doubled $\epsilon$ structure shows that these expressions are symmetric under the interchange of any elements
\be
\epsilon \epsilon A^{d-n} B^n = \epsilon \epsilon  B^n A^{d-n} = \epsilon \epsilon  A B^n A^{d-n-1}  \dots \text{ etc.} \, . 
\ee
This can be generalised to additional tensors, e.g.
\begin{eqnarray}
\epsilon \epsilon A^{d-n-m} B^n C^m &\coloneqq& \epsilon^{\mu_1 \dots \mu_d}\epsilon^{\nu_1 \dots \nu_d} A_{\mu_1 \nu_1} \dots A_{\mu_{d-n-m} \nu_{d-n-m}}\nn \\
&& B_{\mu_{d-n-m+1} \nu_{d-n-m+1}} \dots B_{\mu_{d-m} \nu_{d-m}} C_{\mu_{d-m+1} \nu_{d-m+1}} \dots C_{\mu_{d}\nu_d} \, . 
\end{eqnarray}
The term with $n=0$ in \eqref{gal1} is simply a tadpole term in the Lagrangian $\pi$. It is sometimes neglected, but it contributes to the equations of motion in a manner invariant under the Galileon symmetry. 

Varying the action \eqref{gal1} gives
\be
\delta S_1 =\sum_{n=0}^d  \int \d^d x \[  c_n \delta \pi \, \epsilon \epsilon \eta^{d-n} (\partial \partial \pi )^{n} + n c_n \pi \epsilon \epsilon \eta^{d-n} (\partial \partial \pi )^{n-1} \partial \partial \delta \pi \] \, ,
\ee
which upon twice integrating by parts and using the antisymmetry properties of the double $\epsilon$ structure gives
\be
\delta S_1 =\sum_{n=0}^d  \int \d^d x \[ (1+n)  c_n \delta \pi \, \epsilon \epsilon \eta^{d-n} (\partial \partial \pi )^{n}  \] \, ,
\ee
hence
\be
\frac{\delta S_1}{\delta \pi(x)} = \sum_{n=0}^d(1+n)  c_n  \, \epsilon \epsilon \eta^{d-n} (\partial \partial \pi(x) )^{n}  \, .
\ee
A second form of the same action is
\be \label{gal2}
S_2 = \int \d^d x \[ -\sum_{n=1}^d c_n  \, \epsilon \epsilon (\partial \pi \partial \pi)\eta^{d-n} (\partial \partial \pi )^{n-1} + c_0 \pi  \epsilon \epsilon \eta^d\] \, .
\ee
To obtain this, we split $(\partial \partial \pi )^{n}=(\partial \partial \pi )^{n-1} (\partial \partial \pi)$ for $n\ge 1$ and then integrate by parts once from the last term.
A third form of the action which is in fact more commonly used is
\be
S_3=\int \d^d x \[ \sum_{n=1}^d b_n  \, X \epsilon \epsilon \eta^{d-n+1} (\partial \partial \pi )^{n-1} + c_0 \pi  \epsilon \epsilon \eta^d\] \, ,
\ee
where $X= - \frac{1}{2} \partial_{\mu}\pi \partial^{\mu} \pi$. To demonstrate equivalence, it is easiest to consider its variation
\ba
 \delta S_3 &=& \int \d^d x \[ \sum_{n=1}^d b_n  \, \delta X \epsilon \epsilon \eta^{d-n+1} (\partial \partial \pi )^{n-1} + 
 \sum_{n=1}^d b_n (n-1)  X \epsilon \epsilon \eta^{d-n+1} (\partial \partial \pi )^{n-2}(\partial \partial \delta \pi ) \right. \nn \\
&& \left. + c_0 \delta \pi  \epsilon \epsilon \eta^d\] \, .
\ea
Given $\delta X = -\partial_{\rho} \pi \partial^{\rho} \delta \pi$, 
integrating by parts we have
\ba
 \delta S_3 = \int \d^d x \, \delta \pi &\Bigg[& \sum_{n=1}^d b_n  \, \partial^{\rho} \( \partial_{\rho} \pi \epsilon \epsilon \eta^{d-n+1} (\partial \partial \pi )^{n-1} \) \\
 &+& 
 \sum_{n=1}^d b_n (n-1)  \epsilon \epsilon (\partial \partial X) \eta^{d-n+1} (\partial \partial \pi )^{n-2} + c_0  \epsilon \epsilon \eta^d\Bigg] \, .\nn
\ea
Although not immediately clear, all higher order derivatives cancel due to the following identity \cite{deRham:2010ik} (see Appendix \ref{appendix})
 \be
 \label{eq:deri}
\partial^{\rho} \( \partial_{\rho} \pi \epsilon \epsilon \eta^{d-n+1} (\partial \partial \pi )^{n-1} \) +  (n-1)  \epsilon \epsilon (\partial \partial X) \eta^{d-n+1} (\partial \partial \pi )^{n-2} = (d-n+1) \epsilon \epsilon \eta^{d-n} (\partial \partial \pi )^{n} \, .
 \ee
Hence
\be
\frac{\delta S_3}{\delta \pi(x)} =\sum_{n=1}^d b_n  (d-n+1) \epsilon \epsilon \eta^{d-n} (\partial \partial \pi )^{n}+ c_0 \epsilon \epsilon \eta^{d} \, ,
\ee
so that we may identify $b_n  (d-n+1)=(n+1)c_n$ for $n \ge 1$.

 As a final expression, splitting $(\partial \partial \pi )^{n-1}=(\partial \partial \pi )^{n-2} (\partial \partial \pi)$ for $n\ge 2$ and integrating by parts we have 
 \be
S_4= \int \d^d x \[ -\sum_{n=2}^d b_n  \,  \epsilon \epsilon (\partial \pi \partial X)\eta^{d-n+1} (\partial \partial \pi )^{n-2} + (b_1 X+c_0 \pi)  \epsilon \epsilon \eta^d\] \, .
 \ee
 We shall see that all of the different versions naturally emerge when we consider generalisations.

 \subsection{Differential Form Notation}

The double epsilon structure central to the above arguments is familiar in the Einstein-Cartan formulation of gravity. However, in that case, the first index is a diffeomorphism index, and the second is a local Lorentz index, e.g. in $d=4$
\be
\sqrt{-g} R = \frac{1}{4}\epsilon^{\mu_1 \mu_2 \mu_3 \mu_4}\epsilon_{a_1a_2a_3a_4} e^{a_1}_{\mu_1} e^{a_2}_{\mu_2} R^{a_3 a_4}_{\mu_3 \mu_4} \, .
\ee
In the decoupling limit $\mpl \rightarrow \infty$  the diffeomorphism symmetry is replaced by a global Poincar\'e symmetry of the background Minkowski spacetime, whose vector indices can be identified with the local Lorentz ones (since $e^a_{\mu} \rightarrow \delta_{a \mu}
$), together with linear spin-2 gauge invariance. 

Anticipating the connection with gravitational theories, it proves to be convenient to reexpress the Galileon action in terms of differential forms. For this, we use the analogous shorthand notation for forms
\be
\epsilon A B C .... = \epsilon_{a_1 a_2 a_3 \dots} A^{a_1} \wedge B^{a_2} \wedge C^{a_3} \dots  \, ,
\ee
with one-forms $A^a$ related to the previous two tensors as $A^a = A^a_{\mu} \d x^{\mu}$.
Comparing with the tensor double-$\epsilon$ notation, we have at the integral level
\be
\int \d^d x \epsilon \epsilon A^{d-n} B^n = \int \epsilon A^{d-n} B^n \, ,
\ee
where the LHS is the tensor notation, and the RHS the differential form. As an explicit example, we have
\be
\int \d^4 x\,
\epsilon^{\mu_1\mu_2\mu_3\mu_4}
\epsilon_{a_1a_2a_3a_4}
A^{a_1}_{\mu_1}
B^{a_2}_{\mu_2}
C^{a_3}_{\mu_3}
D^{a_4}_{\mu_4} =\int
\epsilon_{a_1a_2a_3a_4}
A^{a_1}\wedge B^{a_2}\wedge C^{a_3}\wedge D^{a_4}\, .
\ee
In the form notation, the metric is replaced by
\be
\eta_{ \mu a} \d x^\mu = \d x_a \, ,
\ee
and 
$\partial \partial \pi$ is replaced by
\be
\partial_\mu \partial_a \pi \d x^{\mu} = \d \partial_a \pi \, .
\ee
The Galileon action \eqref{gal1} can now be written in this shorthand as
\be \label{gal4}
S_1  =  \int \[ \sum_{n=0}^d c_n \pi \,  \epsilon ( \d x )^{d-n} (\d \partial \pi )^{n} \] \, ,
\ee
so that one $\epsilon$ is explicit, acting on the Lorentz indices, and one is implicit in the wedge products.

 \section{Galileon Duality as  Diffeomorphic Scalar Duality}

\label{GalileonDuality}
\subsection{Diffeomorphism Invariance from Underlying Gravitational Origin}
The Galileon duality transformation was observed to arise in \cite{BiHiguchi,deRham:2013hsa,deRham:2014lqa} in the decoupling limit of massive gravity and bigravity theories as a remnant of the underlying diffeomorphism invariance that arises in the \stu formulation. It had previously been considered in \cite{Curtright:2012gx} from the different perspective of Legendre transformations in the formulation of the action. Subsequently, it was interpreted from the perspective of the coset construction for nonlinearly realised symmetries \cite{Creminelli:2014zxa,Kampf:2014rka,Baratella:2015yya}. 

At a fundamental level, the duality corresponds to a one-parameter, non-local field redefinitions that can be interpreted as a combination of a local field redefinition and a field-dependent diffeomorphism
\ba
&& \tilde \pi(\tilde x) = \pi(x) +\frac{1}{2} \lambda \partial_{\mu} \pi(x)\partial^{\mu} \pi(x)  \, ,\\
&& \tilde x^{\mu} = x^{\mu}+ \lambda \partial^{\mu} \pi(x) \, .
\ea
Since
\be
\frac{\partial}{\partial x^\nu} \tilde \pi(\tilde x) =\partial_{\nu} \pi(x) + \lambda \partial_{\nu} \partial_{\mu} \pi(x)\partial^{\mu} \pi(x) = \frac{\partial \tilde x^{\mu}}{\partial x^{\nu}} \partial_{\mu} \pi(x) \, ,
\ee
we infer that
\be \label{scalarder}
\tilde \partial_{\mu} \tilde \pi = \partial_{\mu} \pi \, ,
\ee
 hence, the duality is invertible in the sense $\lambda  \leftrightarrow - \lambda$, $\pi \leftrightarrow \tilde \pi$
\ba
&& \pi(x) = \tilde \pi(\tilde x)
-\frac{1}{2}\lambda
\tilde\partial_{\mu}\tilde\pi(\tilde x)
\tilde\partial^{\mu}\tilde\pi(\tilde x)
= \tilde\pi(\tilde x)+\lambda \tilde X(\tilde x),\\
&& x^{\mu} = \tilde x^{\mu}
-\lambda\tilde\partial^{\mu}\tilde\pi(\tilde x).
\ea
What is most remarkable about the transformation is that $\partial_{\mu} \tilde \pi$ is a vector under global Poincar\'e transformations but transforms as a scalar, as seen in \eqref{scalarder}, under the diffeomorphism of the duality. This is consistent with its massive gravity origin, where $\pi(x)$ arises from the \stu fields $\phi^a$, which are scalars under diffeomorphisms and vectors under global Lorentz symmetry.

\subsection{$S$-matrix}
\label{sec:Smatrix1}

It is a general theorem in QFT that the $S$-matrix is invariant under local field redefinitions \cite{Chisholm:1961,Kamefuchi:1961,Salam:1970}, or equivalently that the scattering amplitudes can be constructed from any local operator which is a function of the original field \cite{Borchers1960,Epstein1963} provided that the local operator generates single particle states when applied to the vacuum\footnote{For example, at tree-level $\pi(x) + \alpha \pi^2(x)$ is an acceptable local field, but $\pi^2(x)$ is not as it clearly does not generate single particle states.}. By contrast non-local field redefinitions such as $\pi \rightarrow \pi+ \pi \frac{1}{\Box}\pi $ do not preserve the $S$-matrix as they may lead to additional poles picked out by the LSZ formula and may further introduce additional particle states. Furthermore, non-local transformations do not necessarily have a unique inverse which in itself undermines equivalence.
The fact that the Galileon duality transformation is invertible and perturbatively local is why it avoids the usual pitfalls of non-local transformations. In particular, under the local infinitesimal transformation
\be
\delta \pi(x) = \delta \lambda \, \partial_{\mu} \pi(x)\partial^{\mu} \pi(x) \, ,
\ee
the $S$-matrix is necessarily invariant since $\partial_{\mu} \pi(x)\partial^{\mu} \pi(x)$ is a local operator and, to this order, there is no difference between the duality transformation and a genuine local field redefinition 
\be
\pi(x) \rightarrow \pi(x) + \alpha \partial_{\mu} \pi(x)\partial^{\mu} \pi(x) \, .
\ee
At tree-level the equivalence of the $S$-matrix is straightforward, and explicit examples are given in \cite{deRham:2013hsa}.
Beyond tree-level some care needs to be taken since the transformations are nonlinear in the fields, meaning that at the operator level we should carefully define the composite operators. In order to do this, we need to use a regularisation scheme that adapts to the field-dependent diffeomorphisms. Dimensional regularisation is the most suitable since it avoids explicit introduction of a cutoff scale which is not invariant under diffeomorphisms. Furthermore, as we shall show explicitly in section \ref{pathintegralmeasure}, it leaves invariant the path integral measure.

Defining $\hat O(x) \,$ as the composite operator associated classically with $\partial_{\mu} \pi(x)\partial^{\mu} \pi(x) $, so that the operator deformation is
\be
\delta \hat \pi(x) = \delta \lambda \hat O(x) \, ,
\ee
then the $S$-matrix is invariant under infinitesimal transformation provided the LSZ asymptotic condition is satisfied 
\be
\lim_{x^0 \rightarrow \pm \infty}\langle p | (f(x), \hat O(x))_{x^0} | \Omega \rangle =0 \, ,
\ee
where $f(x)$ is a solution of the free field equations of motion and $(f,g)_{x^0}$ is the associated Klein-Gordon inner product at time $x^0$. 

\subsection{Equivalence map}

To see how duality works at the level of the action, it is simplest to work with the differential form notation for the Galileon, not least because the change of coordinates is straightforward. Beginning with the action \eqref{gal4} and rewriting $\pi$ and $x^a$ in terms of $\tilde \pi $ and $\tilde x^a$ we have
\be
S_1  =  \int \[ \sum_{n=0}^d c_n (\tilde \pi+ \lambda \tilde X )\,  \epsilon ( \d \tilde x-\lambda \d \tilde \partial \tilde \pi )^{d-n} (\d \tilde \partial \tilde \pi )^{n} \] \, ,
\ee
and expanding we have
\be
S_1  =  \int \[ \sum_{n=0}^d \sum_{r=0}^{d-n}(-\lambda)^r \frac{(d-n)!}{r!(d-n-r)!}c_n (\tilde \pi+ \lambda \tilde X )\,  \epsilon ( \d \tilde x)^{d-n-r} (\d \tilde \partial \tilde \pi )^{n+r} \] \, .
\ee
As we have already established, the term with $\tilde X$ in front is itself equivalent to a combination of Galileon interactions, and we can integrate by parts to put it in the same form as the first term.
Hence, after a little rearrangement
\be
S_1  =  \int \left[ \sum_{n=0}^d \tilde c_n(\lambda)\,
\tilde\pi \, \epsilon ( \d \tilde x)^{d-n}
(\d \tilde \partial \tilde \pi )^{n} \right] \, ,
\ee
with 
\be
\tilde c_n(\lambda) = \sum_{r=0}^n c_{n-r} (-\lambda)^r \frac{(1+n-r) (d-n+r)!}{(1+n)r! (d-n)!} \, .
\ee
The latter result can be obtained directly at the level of the equations of motion or equivalently the variations, making use of the identity\footnote{This follows since $\delta (\tilde \pi (\tilde x)) = (\delta \tilde \pi)(\tilde x) + \tilde \partial_{\mu} \tilde \pi \delta \tilde x^{\mu} =(\delta \tilde \pi)(\tilde x) + \lambda \partial_{\mu}  \pi  \partial^{\mu}  \delta \pi = \delta \pi(x) + \lambda \partial_{\mu}  \pi  \partial^{\mu}  \delta \pi  $.}
\be
(\delta \tilde \pi)(\tilde x) = \delta \pi(x) \, ,
\ee
so that
\ba
\delta S_1 &=& \int \d^d x  \delta \pi(x) \sum_{n=0}^d(1+n)  c_n  \, \epsilon \epsilon \eta^{d-n} (\partial \partial \pi(x) )^{n} \, , \\
&=& \int \delta \pi(x) \sum_{n=0}^d(1+n)  c_n  \, \epsilon  (\d x)^{d-n} (\d \partial \pi(x) )^{n} \, , \\
&=& \int \delta \tilde \pi(\tilde x) \sum_{n=0}^d(1+n)  c_n  \, \epsilon  (\d \tilde x- \lambda \d \partial \tilde \pi(\tilde x) )^{d-n} (\d \partial \tilde \pi(\tilde x) )^{n} \, , \\
&=& \int \delta \tilde \pi(\tilde x) \sum_{n=0}^d(1+n)  \tilde c_n(\lambda)  \, \epsilon  (\d \tilde x )^{d-n} (\d \partial \tilde \pi(\tilde x) )^{n} \, .
\ea
To reiterate, every Galileon field theory is dual to a distinct Galileon theory, with different Wilson coefficients $c_n \rightarrow \tilde c_n$. Although we have focused our discussion on the privileged class of operators which lead to second order equations of motion, these conclusions apply to any EFT operator. 

 \section{Generalization to Generic Scalar Field Theories}

\label{GenericScalar}

The rather special nature of the Galileon interactions might give the impression that the duality transformations rely heavily on the existence of an underlying nonlinearly realised symmetry. It is that observation that is  emphasised in multiple approaches, including for instance in the coset interpretation \cite{Creminelli:2014zxa,Kampf:2014rka,Baratella:2015yya}. In this section we shall see, however, that the global symmetry plays no role, and it is the interpretation of the duality as a field-dependent diffeomorphism, which is natural in the massive gravity context \cite{deRham:2014lqa,BiHiguchi} that is most important.

Consider now a generic scalar field $\phi(x)$ in Minkowski spacetime with no special properties other than transforming as a scalar under Poincar\'e. We shall again focus for simplicity on the most general interactions that lead to second-order equations of motion, not because the restriction to these is required or allowed by any EFT principle, but simply that they can be written down in closed form and inevitably dominate the long wavelength dynamics. In fact, those are shown to dominate in quite generic situations \cite{Solomon:2017nlh}. 

The key difference from pure Galileons is that the coefficients in the previous formulation can now be promoted to arbitrary functions of $\phi$ {\it and} the kinetic term $X= - \frac{1}{2} (\partial \phi)^2$. In addition, to maintain complete generality at the price of some redundancy, we should consider linear combinations of $S_1$ through $S_4$. The most general action for a single scalar that leads to second-order equations of motion is then \cite{Deffayet:2009wt,Deffayet:2011gz}
\be
\label{general1}
S =\sum_{\substack{n=0,\ldots,d\\ p,q=0,1\\ n+p+q\le d}}\int \left[
F_{npq}(\phi,X)
\epsilon (\d x)^{d-n-p-q} (\d\partial\phi)^n
(\d\phi\partial\phi)^p (\d X\partial\phi)^q \right].
\ee
Note that $p,q$ cannot be greater than unity as a term such as $\epsilon \epsilon (\d \phi \partial \phi)^2 \dots $ vanishes by virtue of the antisymmetry of the Levi-Civita symbols.

\subsection{Generalized Transformation}
The generalisation of the duality is a transformation of the fields and coordinates that does not introduce higher derivatives than those already present in the above action, in other words, it maps the class of theories \eqref{general1} into themselves. We shall also assume that Poincar\'e invariance is manifest in which case the transformation of the coordinates must be of the form
\be \label{req3}
\tilde x^{\mu} = x^{\mu} + G(\phi,X) \partial^{\mu} \phi \, .
\ee
The possibility of a linear term in $\partial^{\mu} X$ is excluded by the requirement that coefficient functions depend only on $\phi$ and $X$ and not higher derivatives of $X$. The transformation of the one-form
\be
\d \tilde x = \d x + G \, \d \partial \phi +G_{,\phi}\, \d \phi \partial \phi+G_{,X}\, \d X \partial \phi \, ,
\ee
only introduces terms already present in \eqref{general1}, which is crucial for the subsequent arguments. The field itself transforms as
\be
\label{eq:field1}
\tilde \phi = \phi + F(\phi,X) \,,
\ee
so that
\be \label{req1}
\d \tilde \phi = \d \phi +F_{,\phi} \d \phi + F_{,X} \d X \, .
\ee
The final requirement is that the derivatives admit a simple transformation
\be \label{req2}
\tilde \partial_{\mu} \tilde \phi = \Omega(\phi, X) \partial_{\mu} \phi \, ,
\ee
so that
\be
\tilde X = -\frac{1}{2} (\tilde \partial \tilde \phi)^2=\Omega^2(\phi, X) X \,.
\ee
Eq.~\ref{req2} mimics a combination of a scalar diffeomorphism and a conformal transformation, with $\Omega$ the appropriate conformal factor.

It is then evident that any function of $\phi$ and $X$ can be rewritten as a function of $\tilde \phi$ and $\tilde X$.
In addition, 
\be
\d \tilde \phi  \tilde \partial \tilde \phi = \Omega(\phi,X) \(  \d \phi \partial \phi +F_{,\phi} \d \phi \partial \phi + F_{,X} \d X \partial \phi \) \, ,
\ee
which again shows that the transformation only introduces terms already present in \eqref{general1}. Furthermore, the transformation is invertible, and the inverse has the same functional form
\ba 
&& \phi = \tilde \phi- F(\phi,X) = \tilde \phi+ \tilde F(\tilde \phi,\tilde X) \, , \label{transformations1}\\
&& \partial_{\mu} \phi = \frac{1}{\Omega(\phi,X)} \tilde \partial_{\mu} \tilde \phi = \tilde \Omega(\tilde \phi, \tilde X)  \tilde \partial_{\mu} \tilde \phi \, , \label{transformations2} \\
&& x^{\mu}= \tilde x^{\mu}- G(\phi,X) \partial^{\mu} \phi = \tilde x^{\mu}- \frac{G(\phi,X)}{\Omega(\phi,X)} \tilde \partial^{\mu} \tilde \phi=\tilde x^{\mu}+ \tilde G(\tilde \phi,\tilde X) \tilde \partial^{\mu} \tilde \phi \, .\label{transformations3}
\ea
From this we infer
\be
F(\phi,X) = -\tilde F(\tilde \phi,\tilde X) \, , \quad \Omega(\phi,X)=\frac{1}{\tilde \Omega(\tilde \phi,\tilde X)} \, , \quad G(\phi,X)=-\frac{\tilde G(\tilde \phi,\tilde X)}{\tilde \Omega(\tilde \phi,\tilde X)} \, .
\ee
This would not have been the case if we had for example allowed in \eqref{req2} term of the form $\partial_{\mu} X$.
Given that
\be
\d \tilde X = \Omega^2 \d X +2 \Omega \Omega_{,X} X \d X +2 \Omega \Omega_{,\phi} X \d \phi \, ,
\ee
the specific combination
\be
(\d \tilde X \tilde \partial \tilde \phi) = \Omega^3 (\d X \partial \phi)  +2 \Omega^2 \Omega_{,X} X (\d X \partial \phi) +2 \Omega^2 \Omega_{,\phi} X (\d \phi \partial \phi) \, 
\ee
which again clearly preserves the form of the action \eqref{general1}.

\subsection*{Local Field Redefinitions}

In its own right, the field transformation \eqref{eq:field1} can be viewed as a local field redefinition. Derivative-dependent local field redefinitions are, however, generally meaningful only perturbatively within an EFT expansion, since they generate an infinite tower of higher-derivative operators that must be included order by order. The subclass with $G=0$, $\Omega=1+F_{,\phi}$, and $F_{,X}=0$ reduces to the usual derivative-independent field redefinition $\phi \to \phi + F(\phi)$. Generic scalar diffeomorphic transformations go beyond this subclass: they include transformations that are perturbatively local but non-perturbatively non-local, while preserving the S-matrix without requiring an enlargement of the operator basis.

\subsection{Integrability Conditions}

\label{Integrability}

For the scalar diffeomorphic transformation $x\to x +G \partial \phi$, $\phi \to \phi +F$, $\p \phi \to \Omega \p \phi$ to make sense, the functions $F, G, \Omega$ cannot be completely free. The consistency of \eqref{req1}, \eqref{req2}, \eqref{req3} requires
\be
\d \tilde \phi = \tilde \partial_{\mu} \tilde \phi(\tilde x) \d \tilde x^{\mu} \,,
\ee
which is the non-trivial differential requirement
\be
\label{eq:condition1}
\d \phi + \d F=\Omega \d \phi - 2 \Omega X \d G- \Omega G \d X   \, .
\ee
Since $\Omega$, $F$ and $G$ are functions of only $\phi$ and $X$ this leads to a system of two-dimensional differential equations
\begin{subequations}
\label{eq:condition1b}
\begin{align}
& F_{,\phi} = \Omega(1-2X G_{,\phi})-1 \, , \\
& F_{,X} = -\Omega(G+2X G_{,X})\,.
\end{align}
\end{subequations}
which require the integrability condition
\ba
\label{eq:integrability1}
\Omega_{,\phi}G-\(\Omega+2X \Omega_{,X}\)G_{,\phi}
+2X \Omega_{,\phi}G_{,X}=-\Omega_{,X}\,.
\ea
\paragraph{Infinitesimal Transformation \& Integrability:}
Let us consider an infinitesimal transformation for which $\Omega = 1 + \lambda \omega$, $G= \lambda g$, $F=\lambda f$, then to first-order in $\lambda$ we have
\be
\omega \d \phi - 2 X \d g - g \d X = \d f \, ,
\ee
which is equivalent to 
\be
\d \tilde f=\d ( f+2 X g)= \omega \d \phi + g \d X \, . 
\ee
The integrability condition for there to exist a $\tilde f=f+2 X g$ is then
\be
\partial_{X}\omega = \partial_{\phi} g \, ,
\ee
which can be solved as
\be
\omega(\phi,X) = \omega_0(\phi) + \int^X_0 \d X' \partial_{\phi} g(\phi,X') \, ,
\ee
with the integral taken at constant $\phi$.
Then $f$ is given by
\be
f(\phi, X)= f_0+\int^\phi_0 \d \phi' \omega_0(\phi')+ \int^X_0 \d X' g(\phi,X') -2 X g(\phi,X)\, .
\ee
Now $f_0$ is just a constant shift of the field, and $\int^\phi_0 \d \phi' \omega_0(\phi')$ is just a standard local field redefinition. Only the last term is genuinely non-trivial since it mixes the field and the kinetic term together. Choosing for simplicity then $f_0=\omega_0(\phi)=0$, we see that the real content of the generalised duality is a free function $g(\phi,X)$ from which 
\be
\omega(\phi,X)=\int^X_0 \d X' \partial_{\phi} g(\phi,X') \, , \quad f(\phi, X)=\int^X_0 \d X' g(\phi,X') -2 X g(\phi,X) \, .
\ee
\paragraph{Nonlinear Transformation:}
Similarly in the nonlinear theory, we can write the condition \eqref{eq:condition1} as
\be
\label{eq:condition2}
\d(F+2 \Omega X G) = (\Omega-1) \d \phi + \Omega G \d X +2 X G\d \Omega  \, . 
\ee
We may regard $\Omega$ as a free function to be specified. $G$ is then fixed by the integrability condition \eqref{eq:integrability1}
\be
\label{eq:integrability2}
\d (\Omega \d \phi + \Omega G \d X +2 X G\d \Omega  )=0 \, ,
\ee
which in turn fixes $F+2 \Omega X G$ and hence $F$. 
Written in terms of the perturbative variables, the integrability condition is
\be
\partial_{\phi} g =\partial_X \omega - \lambda \( \partial_\phi(\omega g)+2 X \partial_X \omega  \partial_{\phi} g-2  (g+X \partial_X g)\partial_{\phi} \omega\) \, .
\ee
As we have seen, the linearised version of this equation is easily solved, and it is similarly straightforward to solve this equation perturbatively to any order in $\lambda$ by iterating this equation. This does not guarantee a global solution for all free functions $\omega/\Omega$ but in practise we only desire a solution over a finite region of field space for which we trust the EFT.

Naturally,  we can equivalently also solve the system by first specifying $G/g$  and finding an iterative solution for $\Omega/\omega$. Although we cannot explicitly write down the general solution, particular exact non-perturbative solutions are easily obtained for specific choices of $\Omega$ as we demonstrate in Sections \ref{Example1} and \ref{Example2}.

\subsection{Duality Horizon}

\label{dualityhorizon}

The invertibility of the transformations (\ref{transformations1}--\ref{transformations3}) requires that the Jacobian 
\be
\frac{\partial \tilde x^\mu}{\partial x^\nu}=\delta^\mu{}_\nu
+\partial_\nu\left(G(\phi,X)\partial^\mu\phi\right) \, ,
\ee
has non-zero determinant, which can be written in differential form notation as
\be \label{detcondition}
\frac{1}{d!}\epsilon (\d x + \d (G \partial \phi))^d \neq 0 \, .
\ee
It is clear that there can be regions in the field space where this is not satisfied, and in analogy with the Gribov problem in gauge theories \cite{Gribov:1977wm}, we may refer to the boundary in the field space at which the determinant vanishes as the {\it duality horizon}.

However, we should remember that all the theories considered here are effective field theories, for which it is generically not meaningful to consider arbitrarily large fields and gradients. The condition \eqref{detcondition} is usually only violated in regions outside of the regime of validity of the EFT. 

In the gravitational context, this can be viewed as a statement that a given coordinate system may not be globally defined. We do not of course regard this as necessarily a problem, the very definition of a manifold presupposes that a coordinate system is valid only over a chart, \ie a finite patch. As we approach the duality horizon, we can transform to a new duality frame to smoothly interpolate the physics. Demanding that \eqref{detcondition} is satisfied everywhere would be like demanding that a given gauge choice in Yang-Mills is valid for all field configurations, something known to be impossible for covariant gauges \cite{Gribov:1977wm}. Anticipating our gravitational discussion, the equivalent statement is that the two different decoupling limits which are dual to each other will have different regimes of validity in field space. This is closely analogous to the situation for nonlinear sigma models 
where a single field parameterisation need not furnish a global chart on the target-space manifold.

As in the Gribov problem, no problem arises in perturbation theory. However, the duality horizon itself may be indicative of interesting non-perturbative physics, just as the Gribov horizon is argued to be related to confinement (for example see \cite{Dudal:2009xh}). We will return to this in more detail in Section~\ref{pathintegralmeasure}.

\paragraph{Well-Posedness:}
Away from the duality horizon, where the transformation is invertible, the duality map defines a local diffeomorphism and therefore preserves the regular codimension-one character/spacetime nature of any hypersurfaces. In particular, this means that a well-defined Cauchy surface in one representation is mapped onto an equally well-defined Cauchy surface in the dual representation. Away from the duality horizon, the dual descriptions always require the precise same amount of initial data, (albeit re-expressed in their respective dual variables). In other words, a hypersurface that provides a well-posed initial-value surface in one representation is mapped onto an equally admissible hypersurface in the dual representation. In concrete examples, it is easy to see that the duality preserves the notion of causality. Specifically, although the speeds of propagation of different species are superficially different, the ratio of speeds is the same \cite{deRham:2014wnv}, so there is no ambiguity in the relative configuration of the propagation cones.

\subsection{Scalar Transformation -- Example of the DBI-Galileon Duality}

\label{Example1}

A simple exact solution to the integrability condition \eqref{eq:integrability1}, \eqref{eq:integrability2} is obtained by imposing $\Omega=1$ ($\omega=0$), since then the condition reduces to
\be
\partial_{\phi} G =0 \, ,
\ee
\ie $G=G(X)$ and $F=F(X)=\int^X_0 G(X') \d X'-2 X G$. 
As an example, consider the special case
\be
G(X) = \lambda \gamma =\lambda \frac{1}{\sqrt{1-2X}} \, ,
\ee
for which 
\be
F(X) = \lambda \( 1-\frac{1}{\sqrt{1-2X}}\)=\lambda(1-\gamma) \, .
\ee
This is exactly the extension of the Galileon duality to the DBI-Galileon theory \cite{deRham:2010eu} as shown in \cite{Chagoya:2016jyn}.

In particular, since $\tilde \partial \tilde \phi = \partial \phi$ we have $\tilde \gamma = \gamma$, and in differential form notation the DBI-Galileon action is 
\be \label{DBI}
S = \int \gamma^{-1} \sum_{n=0}^d C_n \epsilon (\d x)^{d-n} ( \d (\gamma \partial \phi))^{n} \, ,  
\ee
which are the complete set of terms invariant under nonlinearly realised $d+1$-dimensional Poincar\'e transformations (with the $d$-dimensional part linearly realised) that preserve second order equations of motion \cite{deRham:2010eu,Goon:2011sy}.
Explicitly then, the duality transformation is
\ba
&& \tilde \phi = \phi + \lambda (1-\gamma)\, , \\
&& \tilde x^{\mu} = x^{\mu}+ \lambda \gamma \partial^{\mu} \phi \, ,
\ea
and the DBI-Galileon action \eqref{DBI} transforms as
\ba
S&=&\int \tilde\gamma^{-1}
\sum_{n=0}^d C_n
\epsilon
\left(d\tilde x
-
\lambda d(\tilde\gamma\tilde\partial\tilde\phi)
\right)^{d-n}
\left(
d(\tilde\gamma\tilde\partial\tilde\phi)
\right)^n  \, , \\
&=& \int \tilde \gamma^{-1} \sum_{n=0}^d \tilde C_n(\lambda) \epsilon (\d \tilde x)^{d-n} ( \d (\tilde \gamma \tilde \partial \tilde \phi))^{n} \, , 
\ea
with
\be
\tilde C_n(\lambda) = \sum_{r=0}^n (-\lambda)^r  \frac{(d-n+r)!}{r!(d-n)!}C_{n-r} \, .
\ee
Although the DBI-Galileon possesses a nonlinearly realised symmetry, the duality would apply for any function $G(X)$ and Lagrangian coefficients $C_n(X)$, again demonstrating that the nonlinearly realized symmetry (in this case the higher-dimensional Poincar\'e symmetry) is not essential for the realisation of the duality.

\subsection{Conformal Diffeomorphisms -- Explicit Example with a Symmetry Group}

\label{Example2}
\subsubsection{Conformal Factor}

Remarkably, the previous example can be generalised to generic conformal factors $\Omega(\phi,X)$. As a  non-trivial explicit example, we consider   conformal factors which depend only on $X$: $\Omega=\Omega(X)$. In other words, here we are dealing with generic transformations under which the kinetic term transforms as 
\ba
X=-\frac 12 (\p \phi)^2 \ \longrightarrow \ \tilde X =\tilde X(X)=\Omega^2(X)X\,.
\ea
Clearly, for this transformation to make sense, the kinetic term cannot completely disappear, which means that $\Omega^2(X)X\ne $ const, or, in other words,
\ba
\label{eq:meaningful}
\p_X(\Omega^2X)=\Omega(\Omega+2X \Omega_{,X}(X))\ne 0\,.
\ea
In this case $\Omega_{,\phi}=0$, and so long as \eqref{eq:meaningful} is satisfied, the integrability condition \eqref{eq:integrability1} reduces to
\be
-\(\Omega+2X\Omega_{,X}(X)\)G_{,\phi}
=
-\Omega_{,X}(X)\quad \Rightarrow \quad
G_{,\phi}
=
\frac{\Omega_{,X}(X)}{\Omega+2X\Omega_{,X}(X)}\,,
\ee
whose general solution is easily found to be
\be
G(\phi,X) = \frac{\Omega_{,X}}{\Omega+2X\Omega_{,X}}\, \phi+G_0(X)\,,
\ee
where $G_0(X)$ is an arbitrary function. The corresponding function $F$ follows from \eqref{eq:condition1b}
\be
F(\phi,X)=\phi\left(
\frac{\Omega^2}{\Omega+2X\Omega_{,X}}-1
\right)-\int^X \d X'\,
\Omega(X')\left[G_0(X')+2X'G_{0,X'}(X')\right]+F_c\,,
\ee
where $F_c$ is an irrelevant constant shift. Thus, any function $\Omega(X)$ satisfying \eqref{eq:meaningful}
in the patch of interest defines a solution of the
integrability condition.

Without loss of generality,  we can always set $F_c=0$ as this corresponds to a trivial shift of the scalar field, already included in a standard local field redefinition. Moreover, we have already seen that the free function $G_0(X)$ can be absorbed in the previous class of scalar duality \ref{Example1}, so again with little loss of generality we can always set $G_0(X)=0$ and $F_c=0$ when considering the general  conformal factor $\Omega=\Omega(X)$ .

\subsubsection{Symmetry Group}

It is not difficult to find global solutions for real $X$ for which  \eqref{eq:meaningful} is satisfied, or in other words for which $\p_X \tilde X (X) \ne 0$. As a special case, let us take
\ba
\tilde X (X)=\sinh\left(e^\alpha \operatorname{arcsinh}X\right)\equiv   f_\alpha(X)\,,
\ea
or in other words 
\ba
\Omega:=\Omega_\alpha(X)=\left(
\frac{f_\alpha(X)}{X}
\right)^{1/2}\,.
\ea
This map is invertible for all real $X$, with inverse
\be
X=\sinh\left(e^{-\alpha}\operatorname{arcsinh}\tilde X\right)\,.
\ee
Equivalently, the inverse transformation belongs to the same family with
$\alpha\rightarrow-\alpha$. Indeed,
\be
\tilde\Omega_{\alpha}(\tilde X)
=\frac{1}{\Omega_\alpha(X)}
=\Omega_{-\alpha}(\tilde X)\,.
\ee
Since
\be
\frac{\d\tilde X}{\d X}
=\Omega_\alpha\left(\Omega_\alpha+2X\Omega_{\alpha,X}\right)
=\frac{e^\alpha
\cosh\left(e^\alpha \operatorname{arcsinh}X\right)
}{\sqrt{1+X^2}}>0\,,
\ee
the differential equations may be solved for all real $X$
and explicitly (setting $G_0(X)=0$ for simplicity)
\be
G_\alpha(\phi,X)
=\phi\,\frac{1}{2X^2}
\(X-e^{-\alpha} \sqrt{1+X^2}\tanh\(e^\alpha \operatorname{arcsinh}X \)\)\,,
\ee
and
\be
F_\alpha(\phi,X)=\phi\left[
\frac{\sqrt{1+X^2}}{e^\alpha\cosh\left(e^\alpha\operatorname{arcsinh}X\right)}
\left(
\frac{\sinh\left(e^\alpha\operatorname{arcsinh}X\right)}{X}
\right)^{3/2}-1\right]
\,.
\ee
This gives an explicit non-perturbative solution of the integrability condition whose
inverse is obtained by the simple replacement $\alpha\rightarrow-\alpha$ and which is
well defined for all real $X$.

We can write the field transformations in a manner in which the symmetry 
\ba
(\alpha,\phi,X,x^{\mu}) \quad \longleftrightarrow \quad (-\alpha,\tilde \phi,\tilde X,\tilde x^{\mu})
\ea
is manifest as
\ba
&& \operatorname{arcsinh}X(x) = e^{-\alpha} \operatorname{arcsinh} \tilde X(\tilde x) \\
&& \frac{\sqrt{1+ X(x)^2}}{ |X(x)|^{3/2}}  \phi(x)=e^{\alpha}\frac{\sqrt{1+ \tilde X(\tilde x)^2}}{|\tilde X(\tilde x)|^{3/2}} \tilde \phi(\tilde x) \, ,\\
&& x^\mu+\frac12G_\alpha(\phi,X)\p^\mu\phi
=\tilde x^\mu+\frac12G_{-\alpha}(\tilde\phi,\tilde X)\tilde\p^\mu\tilde\phi\,.
\ea

\subsubsection{Closure}
Remarkably, the transformations form a one-parameter group, for which the reflection $\alpha \rightarrow -\alpha$ is the inverse. Taking a second transformation from $(\tilde \phi,\tilde X)$ to $(\check{\phi},\check{X})$ of the form
\ba
&& \operatorname{arcsinh}\tilde X(\tilde x) = e^{-\beta} \operatorname{arcsinh} \check X(\check x) \, ,\\
&& \frac{\sqrt{1+ \tilde X(\tilde x)^2}}{ |\tilde X(\tilde x)|^{3/2}}  \tilde \phi(\tilde x)=e^{\beta}\frac{\sqrt{1+ \check X(\check x)^2}}{|\check X(\check x)|^{3/2}} \check \phi(\check x) \, ,
\ea
then the combined transformation is 
\ba
&& \operatorname{arcsinh}X(x) = e^{-(\alpha+\beta)} \operatorname{arcsinh} \check X(\check x) \, ,\\
&& \frac{\sqrt{1+ X(x)^2}}{ |X(x)|^{3/2}}  \phi(x)=e^{\alpha+\beta}\frac{\sqrt{1+ \check X(\check x)^2}}{|\check X(\check x)|^{3/2}} \check \phi(\check x) \, .
\ea
To complete the argument, we need to confirm that the coordinates transform appropriately:
\be
\check x^{\mu}= \tilde x^{\mu} + G_{\beta}(\tilde \phi,\tilde X) \tilde \partial^{\mu} \tilde \phi(\tilde x) = x^{\mu} + \(G_{\alpha}(\phi,X)+\Omega_{\alpha}(X)  G_{\beta}(\tilde \phi,\tilde X) \) \partial^{\mu} \phi(x) \, .
\ee
The group closes if
\be
G_{\alpha+\beta}(\phi,X)=G_{\alpha}(\phi,X)+\Omega_{\alpha}(X)  G_{\beta}(\tilde \phi,\tilde X) \, ,
\ee
which is confirmed explicitly in Appendix~\ref{app:closure}. We stress again that although the emergence of a global symmetry in this example is an appealing feature, it is not required for the duality, and in general, the inverse duality transformation is not required to take the same functional form. 

\subsection{Dual of a Free Theory}

In the case of the Galileon duality, it was shown in \cite{deRham:2013hsa} that a free scalar theory is dual to a Galileon with specific coefficients $c_n$. In $d=4$, the dual theory is a quintic Galileon. Direct calculation of its scattering amplitudes confirms that the $S$-matrix is trivial despite non-zero contributions from individual Feynman diagrams. It is clear from the above analysis that there is a far more general class of actions, with second order equations of motion, that are nevertheless free theories from the scattering point of view. Let us focus on the duality transformations for which $\Omega$ only depends on $X$. $\Omega(X)$ is a free function subject only to \eqref{eq:meaningful}.
The other two functions are then
\ba
&& G(\phi,X) = \frac{\Omega_{,X}}{\Omega+2X\Omega_{,X}}\, \phi+G_0(X) \\
&& F(\phi,X)=\phi\left(
\frac{\Omega^2}{\Omega+2X\Omega_{,X}}-1
\right)-\int^X \d X'\,
\Omega(X')\left[G_0(X')+2X'G_{0,X'}(X')\right]\,,
\ea
up to an irrelevant constant.
We may begin with a free theory in the $\tilde x$ frame
\be
S_{\rm free} = \int \d^d \tilde x \, \tilde X \, . 
\ee
This then transforms to
\be \label{freetheory}
S_{\rm free} = \int \d^d x \, \det\[ \delta^{\mu}{}_{\nu}+ \partial_{\nu}\( \( \frac{\Omega_{,X}}{\Omega+2X\Omega_{,X}}\, \phi+G_0(X)\) \partial^{\mu} \phi \)\] \Omega^2(X) X \, .
\ee
What is remarkable is that $G_0(X)$ and $\Omega(X)$ are free functions (subject only to \eqref{eq:meaningful}), and so Eq.~\eqref{freetheory} {\it appears} to describe a highly non-trivial interacting theory with potentially arbitrary vertices.

In $d=1$, the Jacobian reduces to a single factor. The $G_0(X)$ term can be removed as part of a total derivative, and so we may set it zero. Then the remaining action reduces to an expression of the form
\be
S_{\rm free} = \int \d t \(A(X) + B(X) \phi \ddot \phi\) \, ,
\ee
with
\ba
A(X)
&=&\frac{\Omega^3X}{\Omega+2X\Omega_{,X}},
\nn\\
B(X)
&=& -\Omega^2X
\left[\frac{\Omega_{,X}}{\Omega+2X\Omega_{,X}}
+2X\partial_X
\left(
\frac{\Omega_{,X}}{\Omega+2X\Omega_{,X}}
\right)
\right] .
\ea
By integrating by parts and discarding the boundary terms we can show that this is the same as a $p(X)$ model
\be
S_{\rm free} = \int \d t \, p(X) \, .
\ee
with 
\be
p(X)=A(X)-\sqrt{2X}
\int^X \frac{\d X'}{\sqrt{2X'}}\,B(X') .
\ee
We see that the duality preserves the overall number of propagating degrees of freedom. The final result is consistent with the fact that a pure $P(X)$ theory has a conserved momentum $p_{\phi}=P_{,X}\sqrt{2X}$, which is a function of $X$, and that momentum is directly related to the conserved momentum of the original free particle theory $p_{\tilde \phi}=\sqrt{2\tilde X}$ via the duality. Conversely, in $d=1$ for any $p(X)$ (within reason), we can find an $\Omega(X)$ to demonstrate equivalence with a free theory. 

In higher dimensions, equivalence between \eqref{freetheory} and a free theory is highly non-trivial and remains satisfied at loop-level. As we shall see below, the duality can also be maintained when coupling to other (matter) fields and gravity.

\section{Coupling to Matter}

\label{Mattercoupling}

So far we have discussed the diffeomorphic scalar duality as a transformation of the scalar sector alone. However, in any physical application the scalar will also be coupled to additional matter fields. In the ordinary Galileon case, it was shown in \cite{deRham:2014lqa} that the duality maps arbitrary local couplings to matter to arbitrary local couplings for dual matter fields. The matter fields themselves transform as expected under the field-dependent diffeomorphisms. The same conclusion continues to hold for the diffeomorphic scalar duality described above for the same reasons. 

Let us denote the collection of matter fields by $\psi_I(x)$, where the index $I$ labels the different species, which may include Lorentz indices for spinning particles. The matter Lagrangian will be taken in the generic form
\be
S_{\rm m}=\int \d^d x \,{\cal L}_{\rm m}
\left(\psi_I,\partial \psi_I,\partial^2 \psi_I,\ldots;\phi,\partial \phi,\partial^2\phi,\ldots;\eta_{\mu\nu}\right) \, ,
\ee
which allows for arbitrary coupling to $\phi$, so that the only assumption is locality. We are currently working with a flat Minkowski metric, including its diff-related variations. Couplings to gravity will be introduced in the following section.
The matter action need not respect the Galileon symmetry, nor need it be restricted to second-order scalar equations of motion. 
Under diffeomorphic duality, the coordinates and the scalar transform as
\be
 \tilde x^\mu = x^\mu+G(\phi,X)\partial^\mu\phi \, , \quad  \tilde \phi(\tilde x) = \phi(x)+F(\phi,X) \, ,
\ee
with
\be
\tilde \partial_\mu \tilde \phi=\Omega(\phi,X)\partial_\mu\phi \, ,
\qquad \tilde X=\Omega^2X \, .
\ee
Bosonic matter fields are transformed as tensors of corresponding rank under the field-dependent diffeomorphisms. Thus, for a matter scalar $\chi$ we define
\be
\tilde \chi(\tilde x)=\chi(x) \, .
\ee
For a vector matter field $A_\mu$ we define
\be
\tilde A_\mu(\tilde x)=\frac{\partial x^\nu}{\partial \tilde x^\mu}
A_\nu(x) \, ,
\ee
and similarly for arbitrary tensor fields. 
The duality therefore acts on the matter sector as an ordinary change of coordinates together with a local redefinition of the scalar $\phi$.
It is useful to introduce the Jacobian matrix
\be
J^\mu{}_\nu
\equiv
\frac{\partial \tilde x^\mu}{\partial x^\nu}
=\delta^\mu{}_\nu+\partial_\nu\left(G(\phi,X)\partial^\mu\phi\right) \, ,
\ee
and its obvious inverse
\be
\tilde J^\mu{}_\nu \equiv
\frac{\partial  x^\mu}{\partial \tilde x^\nu}
=\delta^\mu{}_\nu+\tilde \partial_\nu\left(\tilde G(\tilde \phi,\tilde X)\tilde \partial^\mu \tilde \phi\right) \, .
\ee
We see that the Jacobian depends locally on $\phi$, $X$, and second derivatives of $\phi$. Its inverse is likewise a local function of the same dual quantities. Consequently,
\be
\d^d x=\d^d \tilde x\,
\det\left(\frac{\partial x}{\partial \tilde x}\right)=\d^d \tilde x\, \det[\tilde J] \, ,
\ee
is mapped into a local expression in the dual variables. Implicitly, we are assuming $\det[\tilde J]>0$, \ie field configurations that avoid the duality horizon.
For a matter scalar $\chi$, derivatives transform as
\be
\begin{split}
\partial_\mu \chi(x)
&=
\frac{\partial \tilde x^\nu}{\partial x^\mu}
\tilde \partial_\nu \tilde \chi(\tilde x)
=
J^\nu{}_\mu\,\tilde \partial_\nu\tilde\chi(\tilde x)
\\
&=
(\tilde J^{-1})^\nu{}_\mu\,
\tilde \partial_\nu\tilde\chi(\tilde x) \, .
\end{split}
\ee
Higher derivatives transform similarly. For example,
\be
\begin{split}
\partial_\mu\partial_\nu\chi(x)
&=
J^\rho{}_\mu J^\sigma{}_\nu
\tilde\partial_\rho\tilde\partial_\sigma\tilde\chi
+
(\partial_\mu J^\rho{}_\nu)
\tilde\partial_\rho\tilde\chi
\\
&=
(\tilde J^{-1})^\rho{}_\mu
(\tilde J^{-1})^\sigma{}_\nu
\tilde\partial_\rho\tilde\partial_\sigma\tilde\chi
+
(\tilde J^{-1})^\lambda{}_\mu
\tilde\partial_\lambda
\left[
(\tilde J^{-1})^\rho{}_\nu
\right]
\tilde\partial_\rho\tilde\chi \, .
\end{split}
\ee
Although this generates higher derivatives of the scalar field through derivatives of the Jacobian, the resulting interactions are nevertheless local. 

Thus an arbitrary local matter action is mapped to another local matter action,
\be
S_{\rm m} \left[\psi_I,\phi;\eta_{\mu\nu}\right]
=
\tilde S_{\rm m}\left[\tilde \psi_I,\tilde\phi;\eta_{\mu\nu}\right] \,,
\ee
where
\ba
\tilde S_{\rm m}
&=&
\int \d^d\tilde x \,
\det[\tilde J]
{\cal L}_{\rm m}
\left(
\psi_I(\tilde\psi,\tilde\phi),
\partial\psi_I(\tilde\psi,\tilde\phi),
\ldots;
\phi(\tilde\phi,\tilde X),
\partial\phi(\tilde\phi,\tilde X),
\ldots
\right) \\
&=&
\int \d^d\tilde x \,
\tilde{\cal L}_{\rm m}
\left(
\tilde \psi_I,
\tilde \partial \tilde \psi_I,
\ldots;
\tilde\phi,
\tilde\partial \tilde \phi,
\tilde\partial \tilde\partial  \tilde\phi,
\ldots
\right)\,.
\ea
This expression is local provided the inverse duality map exists. In particular, no inverse powers of $\Box$ or other non-local operators are generated.

A useful example is a matter scalar $\chi$ minimally coupled to the original Minkowski metric,
\be
\label{eq:Schi}
S_\chi
=\int \d^d x \left[-\frac12 \eta^{\mu\nu}\partial_\mu\chi\partial_\nu\chi-V(\chi)
\right] .
\ee
Under the duality, this becomes
\be
\begin{split}
\tilde S_\chi
&=
\int \d^d\tilde x\,
\det[\tilde J]
\left[
-\frac12
\eta^{\mu\nu}
J^\rho{}_\mu J^\sigma{}_\nu
\tilde\partial_\rho\tilde\chi
\tilde\partial_\sigma\tilde\chi
-
V(\tilde\chi)
\right]
\\
&=
\int \d^d\tilde x\,
\det[\tilde J]
\left[
-\frac12
\eta^{\mu\nu}
(\tilde J^{-1})^\rho{}_\mu
(\tilde J^{-1})^\sigma{}_\nu
\tilde\partial_\rho\tilde\chi
\tilde\partial_\sigma\tilde\chi
-
V(\tilde\chi)
\right] .
\end{split}
\ee
Equivalently, the matter field propagates in an effective densitised inverse metric
\be
\begin{split}
\tilde {\cal Z}^{\rho\sigma}
&=
\det[\tilde J]\,
\eta^{\mu\nu}
J^\rho{}_\mu J^\sigma{}_\nu
\\
&=
\det[\tilde J]\,
\eta^{\mu\nu}
(\tilde J^{-1})^\rho{}_\mu
(\tilde J^{-1})^\sigma{}_\nu \, ,
\end{split}
\ee
so that
\be
\tilde S_\chi
=
\int \d^d\tilde x
\left[
-\frac12
\tilde {\cal Z}^{\rho\sigma}
\tilde\partial_\rho\tilde\chi
\tilde\partial_\sigma\tilde\chi
-
\det[\tilde J]\,
V(\tilde\chi)
\right] .
\ee
This is local, but it is no longer minimally coupled to the background Minkowski metric. Instead, the scalar $\tilde \chi$ is coupled to an effective geometry determined by the dual scalar configuration. However, this representation is fully equivalent to the canonical scalar field  \eqref{eq:Schi}. In this case, the equivalent is relatively clear since all that happened is a gauge transformation and $\phi$ is not even dynamical.

The equivalence is non-trivial when the matter couples directly to the field $\phi$ from the outset. For example, a local interaction
\be
S_{\rm int}
=
\int \d^d x \,
{\cal J}(\psi_I,\partial\psi_I,\ldots)\,
H(\phi,X) \, ,
\ee
is mapped to
\be
\tilde S_{\rm int}
=
\int \d^d\tilde x\,
\det[\tilde J]
\tilde{\cal J}(\tilde\psi_I,\tilde\partial\tilde\psi_I,\ldots;\tilde\phi,\tilde X)
H\left(
\tilde\phi+\tilde F(\tilde\phi,\tilde X),
\tilde \Omega(\tilde\phi,\tilde X)^2\tilde X
\right) ,
\ee
where $\tilde{\cal J}$ denotes the transformed local matter operator. 

To complete the discussion for matter, we need to also state how physical fermions transform, \ie how to deal with the spinor indices. For this, it is helpful to remember that when we couple spinors to gravity, the spinor index is associated with the local Lorentz group and not the diffeomorphism group. While the duality corresponds to field-dependent diffeomorphism, there is no accompanying local Lorentz transformation required. Thus, the spinor index does not need to transform, so, for example, for a Dirac fermion field, the appropriate duality transformation can be taken to be ($\alpha$ is spinor index)
\be
\tilde \psi_{\alpha}(\tilde x) =\psi_{\alpha}(x) \, . 
\ee
This perspective will be confirmed by the consideration in the next section where it is argued that the spin-connection is unchanged under the duality, meaning that there is no accompanying local Lorentz transformation. When considering higher spin fermions, we can choose to work in a mixed representation with both spinor and tensor indices, or a pure multispinor notation. The duality transformation is most transparent in the latter case for which for a spin $J/2$ field we need multispinors with $J$ indices
\be
\tilde \psi_{\alpha_1 \dots \alpha_J}(\tilde x) =\psi_{\alpha_1 \dots \alpha_J}(x)  \, .
\ee
In a mixed representation, as is typically used for example for the spin 3/2 Rarita-Schwinger field then we just need to separately consider the transformation of the spinor indices and spacetime indices 
\be
\tilde  \psi_{\alpha \mu}(\tilde x) =\psi_{\alpha \nu}(x) \tilde J^{\nu}{}_{\mu} \, .
\ee

\subsection{Freedom of choice of scalar}

Given that the duality is general and applies to any scalar field, in a theory with multiple scalars we may choose different fields or different combinations of fields to apply to the duality transformation to. Thus, for example, in a theory with two scalar fields $\phi$ and $\chi$, we may choose to apply the duality to each field separately. As in the case of the Galileon duality, what is unusual is the field that is clearly a vector under global transformations effectively behaves like a (conformal) scalar under the field-dependent diffeomorphism. This is illustrated in Table~\ref{tab:Transformations on Minkowski}.

\begin{table}[H]
    \centering
    \begin{tabular}{c|c|c|c}
       Field  & Global Lorentz & $\phi$-Diff & $\chi$-Diff  \\ \hline \hline
        $\phi$ & Scalar & Shift & Scalar\\
 &  & $\phi\to \phi +F(\phi,X)$ & \\ \hline 
        $\p_\mu \phi \leftrightarrow \Phi^a$ & Vector & Conformal Scalar & Vector\\ 
 & & $\Phi^a \to \Omega(\phi,X) \, \Phi^a$ & \\
        \hline
        $\chi$ & Scalar & Scalar & Shift \\
        &  &  & $\chi\to \chi +F(\chi,(\p \chi)^2)$ \\ \hline
        $\p_\mu \chi\leftrightarrow \chi^a$ & Vector & Vector & Conformal Scalar  \\
    &  & & $\chi^a \to \Omega(\chi,(\partial \chi)^2)\, \chi^a$  \\ \hline
        $A_\mu $ & Vector & Vector & Vector \\ \hline 
Spin-J & $(0,J)$- & $(0,J)$- & $(0,J)$-\\
 & Tensor & Tensor & Tensor\\ \hline 
Spin-J/2 & Multispinor & Scalar & Scalar \\ \hline
    \end{tabular}
    \caption{Transformations on Minkowski. By Spin-J/2 we mean a fermion (half-integer spin) written in pure multi-spinor basis $\psi_{\alpha_1 \dots \alpha_{J}}$ and by spin-J, we mean an integer spin written in tensor notation. }
    \label{tab:Transformations on Minkowski}
\end{table}

\section{Coupling to Gravity}

\label{gravity}

The diffeomorphic scalar duality as we have described it so far is the combination of a local field redefinition and a field-dependent diffeomorphism. It is the latter aspect which makes it unusual since the scalar theory is defined in a fixed Minkowski spacetime, but the duality acts non-trivially on the coordinates. If we couple the scalar to gravity, then diffeomorphism invariance becomes a local symmetry, and the coordinates become fundamentally irrelevant, so what, then, is the content of the duality?

\subsection{Duality as a Local Field Redefinition}

The answer lies in the description of Minkowski space. Using vielbeins to describe a general metric $g_{\mu\nu} = \eta_{ab} e_{\mu}^a e_{\nu}^b  $, where $a,b$ are local Lorentz indices, the Minkowski spacetime in Cartesian coordinates is described by the local Lorentz index valued one-form
\be
e^a  = \d x^a \, .
\ee
The duality as described so far maps this to a second one-form which also describes a Minkowski spacetime, but one whose coordinates are non-trivially related
\be \label{viel1}
\tilde e^a = \d \tilde x^a = \d x^a + \d ( G(\phi,\mathcal X) \partial^a \phi ) = e^a +\d ( G(\phi,\mathcal X) \partial^a \phi ) \, .
\ee
In other words, the duality is a field ($\phi$)-dependent local transformation of the vielbein, and hence the metric. To generalise \eqref{viel1} to dynamical gravity, we need to write it in a form for which both vielbeins transform in the conventional manner under diffeomorphisms and local Lorentz transformations. To do this, we replace $\partial^a \phi$ with $\Phi^a$ which is defined by
\be
e^a \Phi_a = \d \phi \, ,
\ee
in other words $\Phi^a = e^{\mu a} \partial_{\mu} \phi$. This ensures that $\Phi^a$ transforms as a scalar under diffeomorphisms but as a vector under local Lorentz transformations. We then define the Lorentz and diffeomorphism scalar 
\be
\label{eq:defX}
\mathcal X = -\frac{1}{2} \eta_{ab} \Phi^a \Phi^b \, .
\ee
On flat space, in the absence of gravity (and therefore to leading order in the decoupling limit), $\X$ and $X=-\frac 12 \eta^{\mu\nu}\p_\mu \phi \p_\nu \phi$ are the same but when coupling to gravity, $\X\ne X$ and the distinction is important. 
Finally, we need to introduce a one-form spin-connection for the local Lorentz symmetry $\omega^{ab}$ and a covariant derivative 
\be
D[\omega] \Phi^a = \d \Phi^a + \omega^{a}{}_{b} \Phi^b \, .
\ee

In a dynamical gravity theory, the duality can then be interpreted as a local field redefinition of the scalar degree of freedom 
\be
\tilde \phi = \phi + F(\phi,\mathcal X) \, ,
\ee
together with a local field redefinition of the vielbein
\be
\tilde e^a = e^a +D[\omega]\(G(\phi,\mathcal X) \Phi^a \) \, ,
\ee
under which the Lorentz vector transforms as a `conformal' diffeomorphism scalar 
\be
\tilde \Phi^a = \Omega(\phi, \mathcal X) \Phi^a \, \Rightarrow \tilde{\mathcal X} =  \Omega^2 \mathcal X \, .
\ee
So defined, the duality clearly preserves the local Lorentz symmetry and diffeomorphism symmetry. This remains true for any spin-connection $\omega^{ab}$. However, we cannot take $\omega^{ab}$ to be the usual torsionless spin-connection since we require the duality to be invertible
\ba
e^a &=& \tilde e^a - D[\omega]\(G(\phi,\mathcal X) \Phi^a \)=\tilde e^a- D[\omega]\(G(\phi,\mathcal X) \Omega(\phi, \mathcal X)^{-1} \tilde \Phi^a \)  \, , \nn \\
&=& \tilde e^a + D[\tilde \omega]\(\tilde G(\tilde \phi,\tilde{\mathcal X}) \tilde \Phi^a \) \, .
\ea
These equations are only consistent if
\be
\tilde \omega^{ab} = \omega^{ab} \, , \quad \tilde G(\tilde \phi,\tilde{\mathcal X})=-G(\phi,\mathcal X) \Omega(\phi, \mathcal X)^{-1} \, .
\ee
By contrast, had we defined $\omega$ and $\tilde \omega$ via the torsion-free conditions $\d e^a + \omega^{a}{}_b \wedge e^b=0$, $\d \tilde e^a + \tilde \omega^{a}{}_b \wedge \tilde e^b=0$ then we would have $\tilde \omega \neq \omega$. The resolution is to work in the {\it first-order} Einstein-Cartan formulation where $\omega$ is regarded as an independent variable in the action, fixed by extremising the action rather than imposing it is torsion-free as a constraint (see \cite{deRham:2015cha} for a similar situation). The first-order formulation allows us to define the duality transformation by the combined local field redefinitions
\be \label{duality10}
 \tilde \phi = \phi + F(\phi,\mathcal X) \, \quad \tilde e^a = e^a +D[\omega]\(G(\phi,\mathcal X) \Phi^a \) \, , \quad \tilde \Phi^a = \Omega(\phi, \mathcal X) \Phi^a \, \quad \tilde \omega^{ab}= \omega^{ab} \, , \nn \\
\ee
with inverse
\ba \label{duality10inverse}
&&  \phi = \tilde \phi + \tilde F(\tilde \phi,\tilde{\mathcal X}) \, \quad e^a = \tilde e^a +D[\tilde \omega]\(\tilde G(\tilde \phi,\tilde{\mathcal X}) \tilde \Phi^a \) \, , \quad  \Phi^a = \tilde \Omega(\tilde \phi, \tilde{\mathcal X}) \tilde \Phi^a \, \quad  \omega^{ab}= \tilde \omega^{ab} \, ,\nn \\
&& \tilde F(\tilde \phi,\tilde{\mathcal X})=-F(\phi,\mathcal X) \, , \, \tilde G(\tilde \phi,\tilde{\mathcal X})=-G(\phi,\mathcal X) \Omega(\phi, \mathcal X)^{-1}  \, , \,  \tilde \Omega(\tilde \phi, \tilde{\mathcal X})=\Omega(\phi, \mathcal X)^{-1}  \, ,
\ea
without contradiction. 

\subsection{Integrability Condition (with Gravity)}

There remains the integrability condition which is the non-trivial requirement that
\be \label{integra1}
\tilde e^a \tilde \Phi_a = \d \tilde \phi \, .
\ee
Using $\mathcal X=-\frac12\Phi_a\Phi^a$ and $D[\omega]\eta_{ab}=0$, one has
\be
\Phi_a D[\omega]\Phi^a
=\frac12 D[\omega](\Phi_a\Phi^a)
=-\d \mathcal X .
\ee
Since $G(\phi,\mathcal X)$ is a Lorentz scalar, $D[\omega]G=\d G$, and hence
\be
\Phi_a D[\omega]\bigl(G\Phi^a\bigr)
=\Phi_a\Phi^a\,\d G
+G\,\Phi_a D[\omega]\Phi^a
=
-2\mathcal X\,\d G-G\,\d \mathcal X .
\ee
Remarkably, then the integrability condition needed to ensure \eqref{integra1} is identical to before
\be
\Omega \d \phi- 2 \Omega \mathcal X \d G- \Omega G \d \mathcal X= \d \phi + \d F \, . 
\ee
In other words, whenever there exists a generalisation of the duality in Minkowski spacetime, this can be immediately extended to a gravitational theory in first-order Einstein-Cartan formulation.

\subsection{Covariant Action}

It remains only to construct the gravitational analogue of the leading low-energy
interactions which map into themselves under the duality. Allowing for arbitrary
powers of the curvature two-form in the wedge-form structure, the natural
extension of \eqref{general1} is
\ba \label{general2}
S[e,\omega,\phi]
&=&
\int
\sum_{r=1}^{\lfloor d/2\rfloor}
\sum_{p,q=0}^{1}
\sum_{m=0}^{d-2r-p-q}
M^{(r)}_{mpq}(\phi,\mathcal X)\,
\epsilon\,
e^{d-2r-m-p-q}
\bigl(D[\omega]\Phi\bigr)^m
\bigl(\d\phi\,\Phi\bigr)^p
\bigl(\d \mathcal X\,\Phi\bigr)^q
R[\omega]^r
\nn \\
&+&
\int
\sum_{p,q=0}^{1}
\sum_{n=0}^{d-p-q}
F_{npq}(\phi,\mathcal X)\,
\epsilon\,
e^{d-n-p-q}
\bigl(D[\omega]\Phi\bigr)^n
\bigl(\d\phi\,\Phi\bigr)^p
\bigl(\d \mathcal X\,\Phi\bigr)^q .
\ea
Here, $R[\omega]^r$ denotes the wedge product of $r$ curvature two-forms,
with Lorentz indices contracted by the Levi-Civita tensor together with the
remaining vielbeins and $\Phi$ insertions. The upper bound
$m\leq d-2r-p-q$ ensures that each term is a $d$-form.  The conventional Einstein-Hilbert kinetic
term arises from
\be
M^{(1)}_{000}=\frac{\mpl^{d-2}}{2(d-2)!}\, .
\ee
In this form, the action may be regarded as the first-order Einstein-Cartan
formulation of a Lovelock-Horndeski-type class of scalar-tensor theories \cite{Lovelock:1971yv,Lovelock:1972vz,Deffayet:2009wt,Deffayet:2009mn,Horndeski:1974wa,Gubitosi:2012hu,Zumalacarregui:2013pma,Deffayet:2013lga}. These may be regarded as the leading interactions in an EFT expansion which will include higher derivative terms. As in the non-gravitational case, these particular subclass of gravitational interactions are special because of their simple differential form structure and will naturally mix into each other under the duality transformation.

 From \eqref{duality10} we infer
\ba
 && e = \tilde e + \tilde G (D[\tilde \omega] \tilde \Phi)+ \frac{\partial \tilde G}{\partial \tilde{\mathcal X}} (\d \tilde{\mathcal X}\tilde \Phi) + \frac{\partial \tilde G}{\partial \tilde \phi} (\d {\tilde  \phi} \tilde \Phi) \, , \\
 && D[\omega] \Phi =\tilde \Omega D[\tilde \omega] \tilde \Phi + \d \tilde \Omega \tilde \Phi = \tilde \Omega (D[\tilde \omega] \tilde \Phi) + \frac{\partial \tilde \Omega}{\partial \tilde{\mathcal X}} (\d \tilde{\mathcal X}\tilde \Phi) + \frac{\partial \tilde \Omega}{\partial \tilde \phi} (\d {\tilde  \phi} \tilde \Phi) \, ,  \\
 && (\d \phi \Phi) = \tilde \Omega (\d \tilde \phi \tilde \Phi) + \tilde \Omega \frac{\partial \tilde F}{\partial \tilde{\mathcal X}} (\d \tilde{\mathcal X}\tilde \Phi) + \tilde \Omega \frac{\partial \tilde F}{\partial \tilde \phi} (\d {\tilde  \phi} \tilde \Phi) \, , \\
 && (\d \mathcal X \Phi) = \tilde \Omega^3 (\d \tilde{\mathcal X} \tilde \Phi) + 2 \tilde \Omega^2 \tilde{\mathcal X} \frac{\partial \tilde \Omega}{\partial \tilde{\mathcal X}} (\d \tilde{\mathcal X}\tilde \Phi) + 2 \tilde \Omega^2 \tilde{\mathcal X} \frac{\partial \tilde \Omega}{\partial \tilde \phi} (\d {\tilde  \phi} \tilde \Phi) \, , \ea
which is to say that the transformations only generate the same 3 combinations $(\d \tilde{\mathcal X} \tilde \Phi)$, $(\d \tilde \phi \tilde \Phi)$, $(D[\tilde \omega] \tilde \Phi)$ up to coefficient functions which depend on $\tilde \phi$ and $\tilde{\mathcal X}$. Hence, the set of duality transformation clearly only generates terms which are already present in the \eqref{general2}
and just acts to transform the coefficients in the manner
\ba
\tilde M_{mpq}^{(r)}
&=&
\sum_{p',q'=0}^1
\sum_{m'=0}^{d-2r-p'-q'}
\alpha^{(r)}_{mpq;m'p'q'}(\tilde\phi,\tilde{\mathcal X})\,
M^{(r)}_{m'p'q'} ,
\nn \\
\tilde F_{npq}
&=&
\sum_{p',q'=0}^1
\sum_{n'=0}^{d-p'-q'}
\beta_{npq;n'p'q'}(\tilde\phi,\tilde{\mathcal X})\,
F_{n'p'q'}\,.
\ea
So, given $R[\omega]=R[\tilde \omega]$, the dual action is 
\ba \label{general2tilde}
S[e,\omega,\phi]&=&\tilde S[\tilde e,\tilde\omega,\tilde\phi] \nn\\
&=&
\int
\sum_{r=1}^{\lfloor d/2\rfloor}
\sum_{p,q=0}^{1}
\sum_{m=0}^{d-2r-p-q}
\tilde M^{(r)}_{mpq}(\tilde\phi,\tilde{\mathcal X})\,
\epsilon\,
\tilde e^{d-2r-m-p-q}
\bigl(D[\tilde\omega]\tilde\Phi\bigr)^m
\bigl(\d\tilde\phi\,\tilde\Phi\bigr)^p
\bigl(\d\tilde{\mathcal X}\,\tilde\Phi\bigr)^q
R[\tilde\omega]^r
\nn \\
&
+&
\int
\sum_{p,q=0}^{1}
\sum_{n=0}^{d-p-q}
\tilde F_{npq}(\tilde\phi,\tilde{\mathcal X})\,
\epsilon\,
\tilde e^{d-n-p-q}
\bigl(D[\tilde\omega]\tilde\Phi\bigr)^n
\bigl(\d\tilde\phi\,\tilde\Phi\bigr)^p
\bigl(\d\tilde{\mathcal X}\,\tilde\Phi\bigr)^q .
\ea
It is important to stress again that in this covariant version of the duality transformation, there is no field-dependent diffeomorphism needed, the above equations are manifestly diffeomorphism invariant. The transformation is now just an invertible  local transformation of the vielbein, spin-connection, and scalar field, which necessarily satisfies the requirements of the equivalence theorem. 

Note that in general, the equation for the spin-connection that follows from \eqref{general2} is not algebraic (unless $M^{(r)}_{mpq}=0$ for $m\neq 0$), and this may suggest the existence of an Ostrogradski ghost due to higher derivatives arising on integrating out the spin-connection. However, we can always choose the magnitude of the coefficients $M^{(r)}_{mpq}$ so that the mass of any potential ghost in general lies well above the cutoff of the EFT, and hence cannot be regarded as a physical instability. Furthermore, if we begin with a theory with the correct number of degrees of freedom (two tensors plus one-scalar), the locality and invertibility of the field redefinition ensures that this remains true after the transformation.

\subsection{Explicit Example}

As a concrete example, let us focus on the case $d=4$ for which the starting theory is just Einstein Gravity coupled to a scalar with a $p(\phi,\mathcal X)$ model, familiar in the case of many cosmological models such as k-inflation and k-essence \cite{Armendariz-Picon:1999hyi,Armendariz-Picon:2000ulo,Babichev:2009ee}. In Einstein-Cartan notation the starting theory is\footnote{To avoid signature issues we take the convention that for Lorentz indices $\epsilon_{0123}=+1$ and for diffeomorphism indices $\epsilon^{0123}=1$ so that $\det(e_{\mu}^a) = \frac{1}{4!}\epsilon_{a_1a_2a_3a_4}\epsilon^{\mu_1 \mu_2 \mu_3 \mu_4} \Pi_{i=1}^4 e^{a_i}_{\mu_i}>0 $.}
\ba
S &=& \int \d^4 x \sqrt{-g} \[ \frac{\mpl^2}{2} R+ p(\phi,\mathcal X) \] \, , \\
&=& \int \epsilon \left[ \frac{\mpl^2}{4} e^2  R[\omega] + \frac{1}{4!}e^4 \,  p(\phi,\mathcal X) \] \, , \nn \\ 
&=&\int \epsilon_{abcd} \left[ \frac{\mpl^2}{4}  e^a \wedge e^b \wedge R^{cd}[\omega] + \frac{1}{4!} e^a \wedge e^b \wedge e^c \wedge e^d  \, p(\phi,\mathcal X) \]\, . 
\ea
Since this is standard Einstein gravity, varying the action with respect to the spin-connection yields the usual torsion-free condition
\be
\d e^a + \omega^a{}_b \wedge e^b =0 \, .
\ee
Now under the duality transformation we have
\be \label{dual1}
S =\int \epsilon \left[ \frac{\mpl^2}{4} \( \tilde e + D[\tilde \omega](\tilde G \tilde \Phi) \)^2  R[\tilde \omega] + \frac{1}{4!} \(\tilde e + D[\tilde \omega](\tilde G \tilde \Phi) \)^4 \tilde p(\tilde \phi, \tilde{\mathcal X})  \] \, ,
\ee
with
\be
\tilde p(\tilde \phi, \tilde{\mathcal X})=p\(\tilde \phi+\tilde F(\tilde \phi, \tilde{\mathcal X}), \tilde \Omega(\tilde \phi,\tilde{\mathcal X})^2 \tilde{\mathcal X}\) \,.
\ee
The spin-connection $\tilde \omega$ is no longer determined by the torsion-free condition in terms of $\tilde e$. At first sight, \eqref{dual1} does not appear to give rise to second-order equations of motion.  However, this apparent higher-derivative structure is degenerate. Since
\eqref{dual1} is obtained by a local invertible field redefinition from
Einstein gravity coupled to a scalar, it cannot introduce additional propagating
degrees of freedom; therefore, the second-order structure of the original formulation is
 hidden rather than lost. In this situation, the second-order structure is  manifested  covariantly. As a result, the same covariant equations govern the dynamics regardless of which initial Cauchy surface is chosen, as long as one stays clear of any duality horizon.

\subsection{Decoupling Limit}

The gravitational extension described above gives a natural covariant completion of the diffeomorphic scalar duality. If we freeze the vielbein $e^a$ to be equal to Minkowski in Cartesian coordinates $\d x^a$ and set the spin connection $\omega$ to zero so that $R(\omega)=0$, the action \eqref{general2} by construction reduces to that of the generic scalar theory on Minkowski \eqref{general1}. However, doing so is not strictly consistent since in general the graviton fluctuations in \eqref{general2} mix with the scalar in a non-trivial way.

To better understand the scalar-graviton mixing, and also to elucidate the emergence of the diffeomorphic scalar duality, it is helpful to define a decoupling limit where the graviton fluctuations are regarded as weak but are not neglected entirely.
To do this, it will be helpful to isolate the usual Einstein-Hilbert term, with its associated $\mpl$ normalisation, and to choose a different scaling for the remaining gravitational couplings. 
Generic interactions in front of the $R[\omega]$ term may lead to operators that are either higher order in derivatives or prevent constraints from being satisfied. As discussed previously, this is not a problem from an EFT point of view as long as the scale associated with these operators is above the cutoff of the EFT (or in other words, if endowed with this cutoff, the EFT still has a non-trivial regime of validity). A suitable reorganisation that will allow this is then
\ba \label{general10}
S[e,\omega,\phi]
&=&
\int
\frac{\mpl^{d-2}}{2(d-2)!}\,
\epsilon\,
\( e+D[\omega]K \) ^{d-2}
R[\omega] .
\nn \\
&+&
\int
\sum_{r=1}^{\lfloor d/2\rfloor}
\sum_{p,q=0}^{1}
\sum_{m=0}^{d-2r-p-q}
{\cal M}^{(r)}_{mpq}(\phi,\mathcal X)\,
\epsilon\,
e^{d-2r-m-p-q}
\bigl(D[\omega]\Phi\bigr)^m
\bigl(\d\phi\,\Phi\bigr)^p
\bigl(\d \mathcal X\,\Phi\bigr)^q
R[\omega]^r
\nn \\
&+&
\int
\sum_{p,q=0}^{1}
\sum_{n=0}^{d-p-q}
F_{npq}(\phi,\mathcal X)\,
\epsilon\,
e^{d-n-p-q}
\bigl(D[\omega]\Phi\bigr)^n
\bigl(\d\phi\,\Phi\bigr)^p
\bigl(\d \mathcal X\,\Phi\bigr)^q \, ,
\ea
where we have defined
\be
K^a(\phi,\mathcal X,\Phi)
\equiv
H(\phi,\mathcal X)\Phi^a \, .
\ee
The special feature of the kinetic terms on the first line is that the coefficient of $R[\omega]$ becomes an exact form in the Minkowski limit, whereas the same is not true of the generic ${\cal M}^{(r)}_{mpq}(\phi,\mathcal X)$ interactions.

In order to take the decoupling limit, we must choose how each coefficient scales with $\mpl$ as $\mpl \rightarrow \infty$. Specifically, we choose
\ba
{\cal M}^{(r)}_{mpq}(\phi,\mathcal X)
&=&
\mpl^{(d-2)/2}m^{(r)}_{mpq}(\phi,\mathcal X)\,,
\ea
and keep $F_{npq}(\phi,\mathcal X)$ finite.
The reason for this choice of scaling will become apparent shortly; however, keeping the usual Einstein-Hilbert term dominant in the limit $\mpl \rightarrow \infty$ ensures that in this limit there will be a Minkowski solution. 

With this in mind, we can perform the following split of the vielbein 
\be
e^a = \d x^a + \frac{1}{\mpl^{(d-2)/2}} h^a \, ,
\ee
where $h^a=h^a{}_\mu \d x^\mu$ is the spin-2 fluctuation, \ie the graviton. 
We may perform a similar decomposition of the dual vielbein
\be \label{tildee}
\tilde e^a = \d \tilde x^a + \frac{1}{\mpl^{(d-2)/2}} \tilde h^a \, .
\ee
Since we work in first-order Einstein-Cartan variables, the spin connection is treated independently and expanded as
\be
\omega^{ab} = \frac{1}{\mpl^{(d-2)/2}} \mu^{ab} \, , 
\ee
which, given the spin-connection does not transform, is identical to its dual version 
\be \label{dualomega}
\tilde \omega^{ab} = \frac{1}{\mpl^{(d-2)/2}} \tilde \mu^{ab} \, , 
\ee
\ie $\tilde \mu^{ab}=\mu^{ab}$.
The curvature two-form is therefore
\be
R^{ab}[\omega]=\frac{1}{\mpl^{(d-2)/2}}\d \mu^{ab}
+\frac{1}{\mpl^{d-2}}\mu^a{}_c\wedge \mu^{cb} \, .
\ee
In the limit $\mpl\rightarrow \infty$, the nonlinear contribution $\mu\wedge\mu\equiv \mu^\bullet{}_c \wedge \mu^{c\ \bullet}$ disappears from all the curvature interaction terms, except the Einstein-Hilbert term. Due to its distinct scaling behaviour, this term retains a single piece that is quadratic in the spin connection, which is required to determine the spin connection. 

At finite $\mpl$ we can still define the relation between the coordinate $x^a$ and the dual coordinate via
\be 
x^a = \tilde x^a + \tilde G(\tilde \phi, \tilde{\mathcal X}) \tilde \Phi^a \, .
\ee
Then the duality map $e^a = \tilde e^a +D[\tilde \omega]\(\tilde G(\tilde \phi,\tilde{\mathcal X}) \tilde \Phi^a \)$ 
can be translated into a map between the graviton $h^a$ and its dual $\tilde h^a$
\be\label{dualgraviton}
h^a = \tilde h^a + \tilde \mu^{a}{}_b  \tilde G(\tilde \phi, \tilde{\mathcal X})  \tilde \Phi^b \, .
\ee
This transformation is similarly invertible
\be
\tilde h^a = h^a +  \mu^{a}{}_b  G( \phi, \mathcal X)  \Phi^b \, .
\ee
The local Lorentz symmetry linearises in the decoupling limit $\mpl \rightarrow \infty$ to the following
\be
\label{linearlocal}
\mu^{ab} \rightarrow \mu^{ab}- \d \lambda^{ab} \, , \quad h^a \rightarrow h^a + \lambda^{a}{}_b \d x^b \, .
\ee
Thus, $\d\mu^{ab}$ is invariant under the linearised local Lorentz transformations.
Consistent with the above definitions, the dual graviton transforms under local Lorentz transformations in the decoupling limit as
\be
 \tilde h^a \rightarrow \tilde h^a + \lambda^{a}{}_b \d \tilde x^b \, .
\ee 
The scalar field $\phi$ is kept fixed in the decoupling limit. On the other hand $\Phi^a$ is determined from 
\be
e^a \Phi_a = \d \phi \,,
\ee
so that schematically
\ba
&& \Phi^a=\partial^a\phi
-\frac{1}{\mpl^{(d-2)/2}}
h^{a}_{\mu}\partial^\mu\phi
+\mathcal{O}\!\left(\mpl^{-(d-2)}\right) \, , \\
&& D[\omega]\Phi^a
=\d\partial^a\phi
+\frac{1}{\mpl^{(d-2)/2}}
\left[
\mu^a{}_b\,\partial^b\phi
-\d\!\left(h^{a}_{\mu}\partial^\mu\phi\right)
\right]
+\mathcal{O}\!\left(\mpl^{-(d-2)}\right) \, , \\
&& K^a
={\cal K}^a
+\frac{1}{\mpl^{(d-2)/2}}\delta K^a
+\mathcal O\!\left(\mpl^{-(d-2)}\right) \, ,
\\
&& e^a+D[\omega]K^a
=\d x^a+\d{\cal K}^a
+\frac{1}{\mpl^{(d-2)/2}}
\left(
h^a+\d\delta K^a+\mu^a{}_b{\cal K}^b
\right)
+\mathcal O\!\left(\mpl^{-(d-2)}\right) \, ,
\ea
where ${\cal K}^a$ is the decoupling limit version of $K^a$, \ie
\be
{\cal K}^a(\phi,X)
\equiv
H(\phi,X)\partial^a \phi\,,
\ee
so the leading terms in the decoupling limit are 
\be
e^a\rightarrow \d x^a,
\qquad
D[\omega]\Phi^a\rightarrow \d\partial^a\phi,
\qquad
{\X\rightarrow X=-\frac12 \partial_\mu\phi\partial^\mu\phi}\qquad {\rm and} \qquad K^a \rightarrow {\cal K}^a\,.
\ee
Assuming that all the $F_{npq}(\phi,X)$ remain finite in the limit
$\mpl\rightarrow\infty$, the third line of \eqref{general10} becomes
\be
S_\phi^{\rm dec}
=
\int
\left[
\sum_{p,q=0}^1
\sum_{n=0}^{d-p-q}
F_{npq}(\phi,X)
\epsilon
(\d x)^{d-n-p-q}
(\d\partial\phi)^n
(\d\phi\,\partial\phi)^p
(\d X\,\partial\phi)^q
\right] .
\ee
This is precisely the generalised scalar theory considered in the flat-space
discussion.

The non-trivial part of the decoupling limit comes from the curvature
couplings, and this is where the graviton-scalar mixing is found. Given our
choice of scalings, the only curvature-coupling term that survives from the second line of \eqref{general10} in the decoupling limit is
\be
S_{\rm mixing}^{\rm dec}
=
\int
\sum_{p,q=0}^{1}
\sum_{m=0}^{d-2-p-q}
m^{(1)}_{mpq}(\phi,X)\,
\epsilon\,
(\d x)^{d-2-m-p-q}
(\d\partial\phi)^m
(\d\phi\,\partial\phi)^p
(\d X\,\partial\phi)^q
\d \mu \, .
\ee
In this form, it is clear that this expression is invariant under the
linearised local Lorentz transformations \eqref{linearlocal}. 

Finally, the scaling of the first line of \eqref{general10} is chosen so that the leading term that would otherwise naively scale as $\mpl^{(d-2)/2}$ is in fact a total derivative, and can be dropped\footnote{As is always the case, the theory can be defined with the appropriate boundary terms such that this procedure is well defined.}.
The remaining finite part that survives in the decoupling limit is therefore the effective kinetic term for the graviton
\ba\label{general11}
S^{\rm kin}
&=&
\int
\frac{1}{2(d-2)!}\,
\epsilon
\Bigg[
(d-2)
\left(\d x+\d{\cal K}\right)^{d-3}
\d h\,\mu
\nn\\
&&
\hspace{2.2cm}
+
(d-2)
\left(\d x+\d{\cal K}\right)^{d-3}
\left(\mu{\cal K}\right)\d\mu
+
\left(\d x+\d{\cal K}\right)^{d-2}
\mu\wedge\mu
\Bigg] .
\ea
We should remember in this compact notation that $(\mu {\cal K})^a=\mu^a{}_b {\cal K}^b$ and $\mu\wedge\mu\equiv \mu^\bullet{}_c \wedge \mu^{c\ \bullet}$.
We can now put everything together to obtain the complete decoupling limit action, in
index-suppressed notation:
\ba \label{general3}
S^{\rm dec}
&=&
\int
\frac{1}{2(d-2)!}\,
\epsilon
\Bigg[
(d-2)
\left(\d x+\d{\cal K}\right)^{d-3}
\d h\,\mu
+
(d-2)
\left(\d x+\d{\cal K}\right)^{d-3}
\left(\mu{\cal K}\right)\d\mu
\nn\\
&&
\hspace{2.8cm}
+
\left(\d x+\d{\cal K}\right)^{d-2}
\mu\wedge\mu
\Bigg]
\nn\\
&+&
\int
\sum_{p,q=0}^{1}
\sum_{m=0}^{d-2-p-q}
m^{(1)}_{mpq}(\phi,X)\,
\epsilon\,
(\d x)^{d-2-m-p-q}
(\d\partial\phi)^m
(\d\phi\,\partial\phi)^p
(\d X\,\partial\phi)^q
\d \mu 
\nn\\
&+&
\int
\left[
\sum_{p,q=0}^{1}
\sum_{n=0}^{d-p-q}
F_{npq}(\phi,X)\,
\epsilon\,
(\d x)^{d-n-p-q}
(\d\partial\phi)^n
(\d\phi\,\partial\phi)^p
(\d X\,\partial\phi)^q
\right]\, .
\ea
The action \eqref{general3} is invariant under the linearised local Lorentz transformations
\eqref{linearlocal}. Varying with respect to the linearised spin connection gives
\ba
0
&=&
\frac{1}{2(d-2)!}
\int
\epsilon
\Bigg[
(d-2)
\left(\d x+\d{\cal K}\right)^{d-3}
\d h\,\delta\mu
+
(d-2)
\left(\d x+\d{\cal K}\right)^{d-3}
\left(\delta\mu\,{\cal K}\right)\d\mu
\nn\\
&+&
(d-2)
\left(\d x+\d{\cal K}\right)^{d-3}
\left(\mu{\cal K}\right)\d(\delta\mu)
+
2
\left(\d x+\d{\cal K}\right)^{d-2}
\mu\,\delta\mu
\Bigg]
+
\int \mathcal J_{ab}\,\delta\mu^{ab} .
\ea
After reintroducing the Lorentz indices and integrating the
$\d(\delta\mu)$ term by parts with the indices displayed, this gives
\ba
0
&=&
\mathcal J_{ab}
+
\frac{1}{2(d-2)!}\Bigg\{
(d-2)\,
\epsilon_{a_1\cdots a_{d-3}cab}\,
\left(\d x+\d{\cal K}\right)^{a_1}
\cdots
\left(\d x+\d{\cal K}\right)^{a_{d-3}}\,
\d h^c
\nn\\
&& +2\,
\epsilon_{a_1\cdots a_{d-2}c[a}\,
\left(\d x+\d{\cal K}\right)^{a_1}
\cdots
\left(\d x+\d{\cal K}\right)^{a_{d-2}}\,
\mu^{c}{}_{b]}
\nn\\
&&
+
(d-2)
\left(
{\cal K}_{b}\,
\epsilon_{a_1\cdots a_{d-3}a cd}
-
{\cal K}_{a}\,
\epsilon_{a_1\cdots a_{d-3}b cd}
\right)
\left(\d x+\d{\cal K}\right)^{a_1}
\cdots
\left(\d x+\d{\cal K}\right)^{a_{d-3}}\,
\d\mu^{cd}
\nn\\
&&
+
(d-2)\,
\epsilon_{a_1\cdots a_{d-3}cab}\,
\left(\d x+\d{\cal K}\right)^{a_1}
\cdots
\left(\d x+\d{\cal K}\right)^{a_{d-3}}\,
\left(
-\d{\cal K}_{d}\,\mu^{cd}
+
{\cal K}_{d}\,\d\mu^{cd}
\right)
\Bigg\} .
\ea
Here antisymmetrisation is defined with unit weight,
\be
X_{[ab]}=\frac12(X_{ab}-X_{ba}) .
\ee
The effective source term from the curvature mixing terms is
\ba
\mathcal J_{ab}
\equiv
(-1)^{d-1}
\d
\Bigg[
\sum_{p,q=0}^{1}
\sum_{m=0}^{d-2-p-q}
m^{(1)}_{mpq}(\phi,X)\,
\epsilon_{\bullet \cdots \bullet ab}\,
(\d x)^{d-2-m-p-q} (\d\partial\phi)^m
(\d\phi\,\partial\phi)^p
(\d X\,\partial\phi)^q
\Bigg] .\notag
\ea
When the total source $\mathcal J_{ab}$ vanishes and ${\cal K}=0$, $\mu$
is determined by the linearised torsion-free constraint. After integrating out
$\mu$, the first line of \eqref{general3} gives the usual quadratic
Fierz--Pauli action for a massless spin-2 field.

Even when ${\cal K}=0$, a non-zero source $\mathcal J_{ab}$ produces an
unavoidable mixing between the spin-2 graviton and the scalar. This mixing can
be removed by a local shift of the linearised spin connection only if
$\mathcal J_{ab}$ has the same local form as the source generated by such a
shift. In general this requires non-trivial tunings of the functions
$m^{(1)}_{mpq}(\phi,X)$.

When ${\cal K}\neq0$, the situation is more restrictive. The shifted kinetic
sector modifies the equation which determines the spin connection, and a local
shift of $\mu$ can remove the scalar-graviton mixing only after imposing
additional tunings on the function $H$ and on the curvature-mixing
coefficients $m^{(1)}_{mpq}(\phi,X)$.

\subsection{Diffeomorphic Scalar Duality as a Gauge Choice}

\label{gaugechoice}

The decoupling limit action \eqref{general3} can be made invariant under nonlinear diffeomorphisms by introducing \stu fields $Y^a(x)$ and replacing $\d x^a \rightarrow  \d Y^a$, and defining
\be
{\cal K}_Y^a
\equiv
H(\phi,X_Y)\partial_Y^a\phi,
\qquad
X_Y=-\frac12\partial_Y^a\phi\,\partial^Y_a\phi .
\ee
This gives the fully diffeomorphism invariant form
\ba \label{general4}
S^{\rm dec}
&=&
\int
\frac{1}{2(d-2)!}\,
\epsilon
\Bigg[
(d-2)
\left(\d Y+\d{\cal K}_Y\right)^{d-3}
\d h\,\mu
+
(d-2)
\left(\d Y+\d{\cal K}_Y\right)^{d-3}
\left(\mu{\cal K}_Y\right)\d\mu
\nn\\
&&
\hspace{2.8cm}
+
\left(\d Y+\d{\cal K}_Y\right)^{d-2}
\mu\wedge\mu
\Bigg]
\nn\\
&+&
\int
\sum_{p,q=0}^{1}
\sum_{m=0}^{d-2-p-q}
m^{(1)}_{mpq}(\phi,X_Y)\,
\epsilon\,
(\d Y)^{d-2-m-p-q}
(\d\partial_Y\phi)^m
(\d\phi\,\partial_Y\phi)^p
(\d X_Y\,\partial_Y\phi)^q
\d\mu
\nn \\
&+&
\int
\left[
\sum_{p,q=0}^{1}
\sum_{n=0}^{d-p-q}
F_{npq}(\phi,X_Y)\,
\epsilon\,
(\d Y)^{d-n-p-q}
(\d\partial_Y\phi)^n
(\d\phi\,\partial_Y\phi)^p
(\d X_Y\,\partial_Y\phi)^q
\right] .
\ea
Choosing the unitary gauge $Y^a=x^a$, we recover the original scalar theory coupled to a conventional massless spin-2 field  \eqref{general3}.
On the other hand, if we choose the gauge 
\be
Y^a=\tilde x^a+\tilde G(\tilde \phi,\tilde X) \tilde \partial^a \tilde \phi \, ,
\ee
where, to avoid confusion, we have denoted the coordinate of the dual gauge slice by $\tilde x$. We also define
\be
\tilde{\cal K}^a
=
H(\tilde\phi+\tilde F,\tilde\Omega^2\tilde X)
\tilde\Omega\,\tilde\partial^a\tilde\phi
\equiv
\tilde H(\tilde\phi,\tilde X)\tilde\partial^a\tilde\phi .
\ee
Under this gauge choice, the shifted kinetic sector is transformed by the replacements
\be
\d Y
\rightarrow
\d\tilde x+\d\!\left(\tilde G\,\tilde\partial\tilde\phi\right),
\qquad
{\cal K}_Y
\rightarrow
\tilde{\cal K},
\qquad
h
\rightarrow
\tilde h+\mu\,\tilde G\,\tilde\partial\tilde\phi .
\ee
The scalar variables transform as
\be
\partial_Y\phi
\rightarrow
\tilde\Omega\,\tilde\partial\tilde\phi,
\qquad
\phi
\rightarrow
\tilde\phi+\tilde F,
\qquad
X_Y
\rightarrow
\tilde\Omega^2\tilde X .
\ee
Here $\tilde \Omega$ and $\tilde F$ are determined by the integrability condition, and the decoupling-limit version of \eqref{dualgraviton} is
\be
h^a = \tilde h^a + \mu^{a}{}_b  \tilde G(\tilde \phi, \tilde X)  \tilde \partial^b \tilde \phi \, ,
\ee
so that we have the dual formulation of the decoupling limit
\begin{small}
\ba \label{general5}
\tilde S^{\rm dec}
&=&
\int
\frac{1}{2(d-2)!}\,
\epsilon
\Bigg[
(d-2)
\left[
\d\tilde x
+
\d\!\left(\tilde G\,\tilde\partial\tilde\phi\right)
+
\d\tilde{\cal K}
\right]^{d-3}
\d\!\left(
\tilde h+\mu\,\tilde G\,\tilde\partial\tilde\phi
\right)
\mu
\nn\\
&&
\hspace{2.8cm}
+
(d-2)
\left[
\d\tilde x
+
\d\!\left(\tilde G\,\tilde\partial\tilde\phi\right)
+
\d\tilde{\cal K}
\right]^{d-3}
\left(\mu\tilde{\cal K}\right)\d\mu
+
\left[
\d\tilde x
+
\d\!\left(\tilde G\,\tilde\partial\tilde\phi\right)
+
\d\tilde{\cal K}
\right]^{d-2}
\mu\wedge\mu
\Bigg]
\nn\\
&+&
\int
\sum_{p,q=0}^{1}
\sum_{m=0}^{d-2-p-q}
m^{(1)}_{mpq}
\left(
\tilde\phi+\tilde F,
\tilde\Omega^2\tilde X
\right)
\epsilon
\left[
\d\tilde x
+
\d\!\left(\tilde G\,\tilde\partial\tilde\phi\right)
\right]^{d-2-m-p-q}
\left[
\d\!\left(
\tilde\Omega\,\tilde\partial\tilde\phi
\right)
\right]^m
\nn\\
&&
\left[
\tilde\Omega
\left(
\left(
1+\frac{\partial\tilde F}{\partial\tilde\phi}
\right)
(\d\tilde\phi\,\tilde\partial\tilde\phi)
+
\frac{\partial\tilde F}{\partial\tilde X}
(\d\tilde X\,\tilde\partial\tilde\phi)
\right)
\right]^p
\left[
\tilde\Omega
\left(
2\tilde\Omega\tilde X
\frac{\partial\tilde\Omega}{\partial\tilde\phi}
(\d\tilde\phi\,\tilde\partial\tilde\phi)
+
\left(
\tilde\Omega^2
+
2\tilde\Omega\tilde X
\frac{\partial\tilde\Omega}{\partial\tilde X}
\right)
(\d\tilde X\,\tilde\partial\tilde\phi)
\right)
\right]^q
\d\mu
\nn\\
&+&
\int
\sum_{p,q=0}^{1}
\sum_{n=0}^{d-p-q}
F_{npq}
\left(
\tilde\phi+\tilde F,
\tilde\Omega^2\tilde X
\right)
\epsilon
\left[
\d\tilde x
+
\d\!\left(\tilde G\,\tilde\partial\tilde\phi\right)
\right]^{d-n-p-q}
\left[
\d\!\left(
\tilde\Omega\,\tilde\partial\tilde\phi
\right)
\right]^n
\\
&&
\left[
\tilde\Omega
\left(
\left(
1+\frac{\partial\tilde F}{\partial\tilde\phi}
\right)
(\d\tilde\phi\,\tilde\partial\tilde\phi)
+
\frac{\partial\tilde F}{\partial\tilde X}
(\d\tilde X\,\tilde\partial\tilde\phi)
\right)
\right]^p
\left[
\tilde\Omega
\left(
2\tilde\Omega\tilde X
\frac{\partial\tilde\Omega}{\partial\tilde\phi}
(\d\tilde\phi\,\tilde\partial\tilde\phi)
+
\left(
\tilde\Omega^2
+
2\tilde\Omega\tilde X
\frac{\partial\tilde\Omega}{\partial\tilde X}
\right)
(\d\tilde X\,\tilde\partial\tilde\phi)
\right)
\right]^q . \nn
\ea
\end{small}
By reorganising the scalar-dependent contributions originating from the shifted
Einstein--Hilbert kinetic sector
into transformed coefficient functions, the result can be recast in a form analogous to the original decoupling limit. Equivalently, this form can be obtained by taking the decoupling
limit of the dual gravitational action \eqref{general2tilde} directly, making
use of \eqref{tildee} and \eqref{dualomega} and 
using
\be
\d\tilde x+\d\!\left(\tilde G\,\tilde\partial\tilde\phi\right)+\d\tilde{\cal K}
=
\d\tilde x+\d\hat{\cal K},
\qquad
\hat{\cal K}^a=(\tilde H+\tilde G)\tilde\partial^a\tilde\phi \,.
\ee
Since the spin connection is invariant under the duality, we may set
$\tilde\mu=\mu$.
Putting this together we have
\ba \label{general3tilde}
\tilde S^{\rm dec}
&=&
\int
\frac{1}{2(d-2)!}\,
\epsilon
\Bigg[
(d-2)
\left(\d\tilde x+\d\hat{\cal K}\right)^{d-3}
\d\tilde h\,\tilde\mu
+
(d-2)
\left(\d\tilde x+\d\hat{\cal K}\right)^{d-3}
\left(\tilde\mu\hat{\cal K}\right)\d\tilde\mu
\nn\\
&&
\hspace{2.8cm}
+
\left(\d\tilde x+\d\hat{\cal K}\right)^{d-2}
\tilde\mu\wedge\tilde\mu
\Bigg]
\nn\\
&+&
\int
\sum_{p,q=0}^{1}
\sum_{m=0}^{d-2-p-q}
\tilde m^{(1)}_{mpq}(\tilde\phi,\tilde X)\,
\epsilon\,
(\d\tilde x)^{d-2-m-p-q}
(\d\tilde\partial\tilde\phi)^m
(\d\tilde\phi\,\tilde\partial\tilde\phi)^p
(\d\tilde X\,\tilde\partial\tilde\phi)^q
\d\tilde\mu
\nn \\
&+&
\int
\left[
\sum_{p,q=0}^{1}
\sum_{n=0}^{d-p-q}
\tilde F_{npq}(\tilde\phi,\tilde X)\,
\epsilon\,
(\d\tilde x)^{d-n-p-q}
(\d\tilde\partial\tilde\phi)^n
(\d\tilde\phi\,\tilde\partial\tilde\phi)^p
(\d\tilde X\,\tilde\partial\tilde\phi)^q
\right] .
\ea
The $\tilde G$-dependent terms generated by
$\d(\mu\tilde G\tilde\partial\tilde\phi)$ combine with the
$(\mu\tilde{\cal K})\d\mu$ term to give the shifted coefficient
$\hat{\cal K}$ in the dual-frame kinetic sector.
Here, $\tilde m^{(1)}_{mpq}$ and $\tilde F_{npq}$ denote the transformed
coefficient functions obtained from the same duality map as in the full
covariant theory.

\subsection{$p(\phi,\X)$ Decoupling Limit}

Let us now return to the explicit example of Einstein gravity coupled to a scalar
with Lagrangian $p(\phi,\mathcal X)$, and take its decoupling limit in order to see
directly how the diffeomorphic scalar duality acts. For simplicity, we focus on $d=4$. We take
\be
S
=
\int \epsilon
\left[
\frac{\mpl^2}{4} e^2 R[\omega]
+
\frac{1}{4!}e^4 p(\phi,\mathcal X)
\right] \,  .
\ee
In this case, the natural choice is to keep $p(\phi,\mathcal X)$ fixed in the limit $\mpl\rightarrow\infty$, in the sense that $p(\phi,\mathcal X)\to p(\phi,X)$ as $\mpl\to \infty$ and as before, we  express the vielbein and connection as
\be
e^a=\d x^a+\frac{1}{\mpl}h^a,
\qquad
\omega^{ab}=\frac{1}{\mpl}\mu^{ab},
\ee
then the leading term in the Einstein-Hilbert action is proportional to
$\mpl\,\epsilon(\d x)^2\d\mu$, and is a total derivative. The finite part is
therefore
\be
S_{\rm EH}^{\rm dec}
=
\int
\epsilon
\left[
\frac12
\d x\wedge\d h\wedge\mu
+
\frac14
(\d x)^2\wedge\mu\wedge\mu
\right] .
\ee
This is the standard Fierz-Pauli kinetic term (usually expressed with the Lichnerowicz operator) written in Einstein-Cartan form.
Since $p(\phi,\mathcal X)$ is not scaled with $\mpl$, the decoupling limit of the matter sector is simply
\be
S_{\phi}^{\rm dec}
=
\int \frac{1}{4!} \epsilon (\d x)^4 p(\phi,X),
\qquad X=-\frac12\partial_\mu\phi\partial^\mu\phi .
\ee
Thus, the decoupling limit of the original theory is
\be \label{explicitdec1}
S^{\rm dec} =
\int \epsilon \left[ \frac12
\d x\wedge\d h\wedge\mu
+ \frac14
(\d x)^2\wedge\mu\wedge\mu
+ \frac{1}{4!}
(\d x)^4p(\phi,X)
\right] .
\ee
It is self-evident that we are here dealing with a fully decoupled limit, in the sense that the scalar and spin-2 sectors fully decouple. Since the spin-2 is linear, the only non-trivial scattering amplitudes in this limit are those in the scalar sector, and they receive no contributions from graviton exchange at this decoupling limit. 
Varying with respect to $\mu$ gives the
linearised torsion-free condition
\be
\d h+\mu\wedge\d x=0 \, .
\ee
We now take the decoupling limit of the dual action which is obtained from a local field redefinition 
\be
S
=
\int \epsilon
\left[
\frac{\mpl^2}{4}
\left(
\tilde e+D[\tilde\omega](\tilde G\tilde\Phi)
\right)^2
R[\tilde\omega]
+
\frac{1}{4!}
\left(
\tilde e+D[\tilde\omega](\tilde G\tilde\Phi)
\right)^4
\tilde p(\tilde\phi,\tilde{\mathcal X})
\right] \, ,
\ee
where
\be
\tilde p(\tilde\phi,\tilde{\mathcal X})
=
p\left(
\tilde\phi+\tilde F(\tilde\phi,\tilde{\mathcal X}),
\tilde \Omega(\tilde\phi,\tilde{\mathcal X})^2\tilde{\mathcal X}
\right).
\ee
We expand the dual variables as
\be
\tilde e^a=\d\tilde x^a+\frac{1}{\mpl}\tilde h^a,
\qquad
\tilde\omega^{ab}=\frac{1}{\mpl}\tilde\mu^{ab}.
\ee
Using
\be
D[\tilde\omega]
\left(
\tilde G\tilde\Phi^a
\right)
=
\d\!\left(
\tilde G\,\tilde\partial^a\tilde\phi
\right)
+
\frac{1}{\mpl}
\tilde\mu^a{}_b\,
\tilde G\,\tilde\partial^b\tilde\phi
+
\mathcal O(\mpl^{-2}),
\ee
we have
\be
\tilde e^a+D[\tilde\omega](\tilde G\tilde\Phi^a)
=
\d\tilde x^a
+
\d\!\left(
\tilde G\,\tilde\partial^a\tilde\phi
\right)
+
\frac{1}{\mpl}
\left(
\tilde h^a
+
\tilde\mu^a{}_b\,
\tilde G\,\tilde\partial^b\tilde\phi
\right)
+
\mathcal O(\mpl^{-2}).
\ee
The leading term in the dual Einstein-Hilbert sector is again a total
derivative, now proportional to
\be
\mpl\,
\epsilon
\left[
\d\tilde x
+
\d\!\left(
\tilde G\,\tilde\partial\tilde\phi
\right)
\right]^2
\d\tilde\mu \, .
\ee
Dropping this boundary term, the finite dual decoupling limit is
\ba \label{explicitdec2}
\tilde S^{\rm dec}
&=&
\int
\epsilon
\left[
\frac12
\left[
\d\tilde x
+
\d\!\left(
\tilde G\,\tilde\partial\tilde\phi
\right)
\right]
\wedge
\d\!\left(
\tilde h
+
\tilde\mu\,\tilde G\,\tilde\partial\tilde\phi
\right)
\wedge\tilde\mu
\right.
\nn\\
&&
\left.
+
\frac14
\left[
\d\tilde x
+
\d\!\left(
\tilde G\,\tilde\partial\tilde\phi
\right)
\right]^2
\wedge\tilde\mu\wedge\tilde\mu
+
\frac{1}{4!}
\left[
\d\tilde x
+
\d\!\left(
\tilde G\,\tilde\partial\tilde\phi
\right)
\right]^4
\tilde p(\tilde\phi,\tilde X)
\right]  \, .
\ea
This is the explicit decoupling limit of the dual gravitational theory. Varying
with respect to the linearised spin connection gives
\ba
0
=
\int &\Bigg[& \frac12 \epsilon_{abcd}
\left[\d\tilde x^a
+
\d\!\left(
\tilde G\,\tilde\partial^a\tilde\phi
\right)\right]
\wedge
\d\tilde h^b
\wedge
\delta\tilde\mu^{cd}
\nn\\
&+&
\frac12
\epsilon_{abcd}
\left[\d\tilde x^a
+
\d\!\left(
\tilde G\,\tilde\partial^a\tilde\phi
\right)
\right]
\wedge
\left(\delta\tilde\mu^b{}_{e}
\,\tilde G\,\tilde\partial^e\tilde\phi \right)
\wedge
\d\tilde\mu^{cd}
\nn\\
&+&
\frac12
\epsilon_{abcd}
\left[
\d\tilde x^a
+
\d\!\left(
\tilde G\,\tilde\partial^a\tilde\phi
\right)
\right]
\wedge
\left(
\tilde\mu^b{}_{e}
\,\tilde G\,\tilde\partial^e\tilde\phi
\right)
\wedge
\d\!\left(\delta\tilde\mu^{cd}\right)
\nn\\
&+&
\frac12
\epsilon_{abcd}
\left[
\d\tilde x^a
+
\d\!\left(
\tilde G\,\tilde\partial^a\tilde\phi
\right)
\right]
\wedge
\left[
\d\tilde x^b
+
\d\!\left(
\tilde G\,\tilde\partial^b\tilde\phi
\right)
\right]
\wedge
\tilde\mu^c{}_{e}
\wedge
\delta\tilde\mu^{ed}
\Bigg] \, ,
\ea
which corresponds to 
\ba
0
&=&
\frac12
\epsilon_{pqab}
\left[
\d\tilde x^p
+
\d\!\left(
\tilde G\,\tilde\partial^p\tilde\phi
\right)
\right]
\wedge
\d\tilde h^q
\nn\\
&+&
\frac12
\epsilon_{pqab}
\left[
\d\tilde x^p
+
\d\!\left(
\tilde G\,\tilde\partial^p\tilde\phi
\right)
\right]
\wedge
\d\!\left(
\tilde\mu^q{}_{e}
\,\tilde G\,\tilde\partial^e\tilde\phi
\right)
\nn\\
&+&
\frac12
\epsilon_{p[a|cd}
\left[
\d\tilde x^p
+
\d\!\left(
\tilde G\,\tilde\partial^p\tilde\phi
\right)
\right]
\wedge
\tilde G\,\tilde\partial_{|b]}\tilde\phi\,
\d\tilde\mu^{cd}
\nn\\
&+&
\frac14
\left(
\epsilon_{pqrb}
\left[
\d\tilde x^p
+
\d\!\left(
\tilde G\,\tilde\partial^p\tilde\phi
\right)
\right]
\wedge
\left[
\d\tilde x^q
+
\d\!\left(
\tilde G\,\tilde\partial^q\tilde\phi
\right)
\right]
\wedge
\tilde\mu^r{}_{a}
\right.
\nn\\
&&
\left.
\hspace{0.2cm}
-
\epsilon_{pqra}
\left[
\d\tilde x^p
+
\d\!\left(
\tilde G\,\tilde\partial^p\tilde\phi
\right)
\right]
\wedge
\left[
\d\tilde x^q
+
\d\!\left(
\tilde G\,\tilde\partial^q\tilde\phi
\right)
\right]
\wedge
\tilde\mu^r{}_{b}
\right) .
\ea
From the perspective of the dual vielbein and spin-connection, this corresponds to a non-zero torsion for the linearised dual vielbein in the sense
\be
\d\tilde h
+
\tilde\mu
\wedge \d\tilde x  \neq 0 \, .
\ee
The key now is to show that the two decoupling limit theories
\eqref{explicitdec1} and \eqref{explicitdec2}, which no longer carry the
nonlinear diffeomorphism symmetry, are in fact dual to each other under the
diffeomorphic scalar duality. This is defined by the usual non-gravitational transformations
\ba
&&
x^{\mu}
=
\tilde x^{\mu}
+
\tilde G(\tilde\phi(\tilde x),\tilde X(\tilde x))
\tilde\partial^{\mu}\tilde\phi(\tilde x),
\nn\\
&&
\phi(x)
=
\tilde\phi(\tilde x)
+
\tilde F(\tilde\phi(\tilde x),\tilde X(\tilde x)),
\qquad
X(x)
=
\tilde \Omega(\tilde\phi(\tilde x),\tilde X(\tilde x))^2
\tilde X(\tilde x).
\ea
supplemented by the transformation for the graviton
\be
h = \tilde h+ \tilde\mu \,\tilde G\,\tilde\partial\tilde\phi \, .
\ee
Substituting these transformations into \eqref{explicitdec1} gives directly
\ba
S^{\rm dec}
&=&
\int
\epsilon
\left[
\frac12
\left[
\d\tilde x
+
\d\!\left(
\tilde G\,\tilde\partial\tilde\phi
\right)
\right]
\wedge
\d\!\left(
\tilde h
+
\tilde\mu\,\tilde G\,\tilde\partial\tilde\phi
\right)
\wedge
\tilde\mu
\right.
\nn\\
&&
\left.
+
\frac14
\left[
\d\tilde x
+
\d\!\left(
\tilde G\,\tilde\partial\tilde\phi
\right)
\right]^2
\wedge
\tilde\mu\wedge\tilde\mu
+
\frac{1}{4!}
\left[
\d\tilde x
+
\d\!\left(
\tilde G\,\tilde\partial\tilde\phi
\right)
\right]^4
\tilde p(\tilde\phi,\tilde X)
\right].
\ea
This is precisely \eqref{explicitdec2}. This is not a surprise, it is confirming that the diffeomorphic scalar duality is the decoupling limit remnant of the field redefinition/gauge choice made in the gravitational theory. Alternatively, we could state that the decoupling limit commutes with the diffeomorphic scalar duality.

\subsection{Inequivalent Covariantisations}

\begin{figure}
    \centering
    \includegraphics[width=0.6\linewidth]{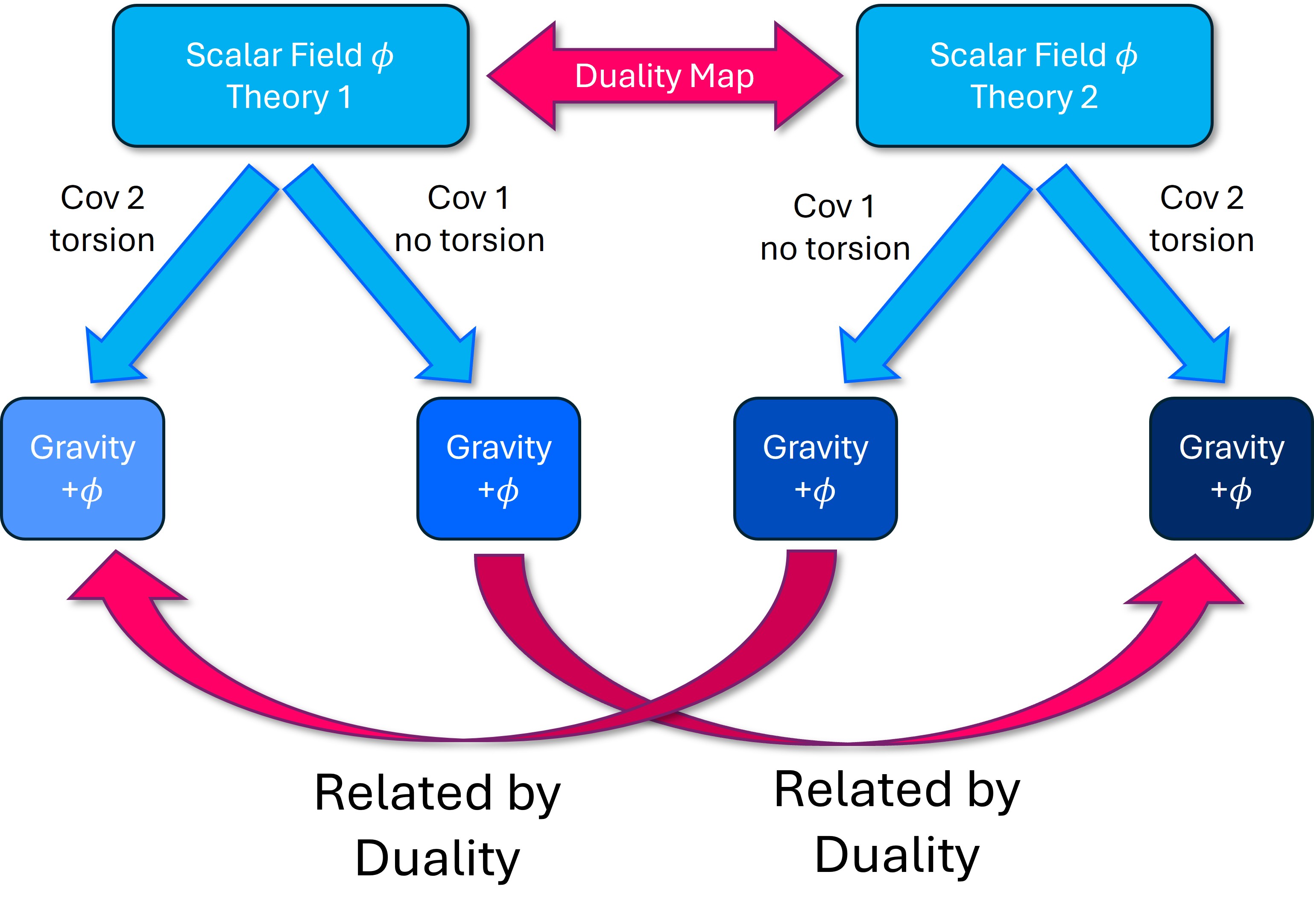}
    \caption{A given scalar field theory in Minkowski can be covariantised, \ie coupled to gravity in multiple ways. The choice of minimal coupling to an Einstein-Hilbert term (no torsion) maps under the duality transformation to a non-minimal kinetic term with non-zero torsion.}
    \label{fig:placeholder2}
\end{figure}

In the previous example, if we discard the graviton perturbations entirely,
\ie set $h=\tilde h=\mu=\tilde \mu =0$, then we recover the pure
Minkowski-space result that
\be \label{Spphi}
S_{\phi}=\int \epsilon \frac{1}{4!}
(\d x)^4p(\phi,X) \, ,
\ee
is dual to 
\be \label{Spphidual}
\tilde S_{\phi}
=\int \epsilon 
\frac{1}{4!}
\left[ \d\tilde x
+
\d\!\left(
\tilde G\,\tilde\partial\tilde\phi
\right) \right]^4
\tilde p(\tilde\phi,\tilde X) \, ,
\ee
under the diffeomorphic scalar duality
\ba
&&
x^{\mu}=\tilde x^{\mu}
+\tilde G(\tilde\phi(\tilde x),\tilde X(\tilde x))
\tilde\partial^{\mu}\tilde\phi(\tilde x),
\nn\\
&& \phi(x)
=
\tilde\phi(\tilde x)+\tilde F(\tilde\phi(\tilde x),\tilde X(\tilde x)),
\qquad
X(x)=\tilde \Omega(\tilde\phi(\tilde x),\tilde X(\tilde x))^2
\tilde X(\tilde x).
\ea
Beginning with the dual theory \eqref{Spphidual} we could choose to add a graviton with no-mixing term
\ba
S^{\rm dec}
&=&\int  \epsilon
\left[\frac12
\d \tilde x\wedge\d \tilde h\wedge\tilde\mu
+\frac14
(\d \tilde x)^2\wedge\tilde\mu\wedge\tilde\mu
\right] + \int \epsilon
\frac{1}{4!}
\left[\d\tilde x
+\d\!\left(
\tilde G\,\tilde\partial\tilde\phi
\right)
\right]^4
\tilde p(\tilde\phi,\tilde X)
\ea
which is the decoupling limit of the following covariant theory 
\be
S
=
\int \epsilon
\left[
\frac{\mpl^2}{4} \tilde e^2 R[\tilde \omega]
+
\frac{1}{4!}(\tilde e+ D[\tilde \omega] (\tilde G \tilde \Phi))^4 \tilde p(\tilde \phi,\tilde{\mathcal X})
\right] \,  .
\ee
Now, performing the inverse duality transformation, we find 
\be
S = \int \epsilon
\left[\frac{\mpl^2}{4} ( e+ D[ \omega] ( G \Phi) )^2R[\omega] + \frac{1}{4!}e^4 p(\phi,\mathcal X) \right] \, ,
\ee
which corresponds to the original scalar theory non-minimally coupled to gravity in such a way that there is a non-zero torsion in this frame 
\be
\d e + \omega \wedge e \neq 0 \, .
\ee
This illustrates the inherent ambiguity in the covariantisation. Only one duality frame can be viewed as the one in which matter is minimally coupled to gravity with only an Einstein-Hilbert kinetic term, with the usual torsion free requirement. All other dual frames will then inherit non-minimal kinetic interactions. This generic feature is summarised in Figure~\ref{fig:placeholder2}.

\section{Path Integral Equivalence}

\label{Smatrix}

\subsection{Scattering amplitudes}

Let us now turn to the question of how scattering amplitudes behave under the duality transformation. For this it is helpful to view everything from the perspective of the gravitational theory.
In the
covariant Einstein-Cartan formulation, the duality transformation is a local, invertible change of variables,
\be
(e^a,\omega^{ab},\phi,\chi)
\quad\longleftrightarrow\quad
(\tilde e^a,\tilde\omega^{ab},\tilde\phi,\tilde \chi),
\ee
with the spin connection treated as an independent field, and $\chi$ is a stand in for the matter fields. 
When computing Minkowski spacetime scattering amplitudes, we assume the existence of a Minkowski vacuum and perturb accordingly. 
Around an
asymptotically flat background, this change of variables maps the one-particle
asymptotic states into themselves, up to field-basis rotations and the usual
wavefunction normalisations. It therefore falls within the scope of the
equivalence theorem for field redefinitions
\cite{Kamefuchi:1961,Chisholm:1961,Salam:1970,Borchers1960}\footnote{Of course, the original formulations of these equivalence theorems were not phrased with gravitational scattering amplitudes in mind, but such amplitudes are by now routinely computed and are known to explicitly satisfy equivalence under local field redefinitions.}. The tree-level
S-matrix computed in the two covariant descriptions is consequently unchanged.

We can understand what happens in the scalar theory by taking the decoupling limit
$\mpl\rightarrow\infty$. In this limit the graviton interactions contributing
to purely scalar exchange amplitudes decouple, at least in the case where any graviton-scalar mixing terms can be removed with a local field redefinition. Then the scalar amplitudes inherit
the tree-level invariance of the S-matrix, even though the duality now takes the form of a non-local field redefinition. As noted in \cite{deRham:2013hsa,deRham:2014lqa} this is because the infinitesimal form of the duality transformation is a local field redefinition, and if we view the finite transformation as an integrated version of an infinitesimal one, this continues to hold (at least trivially within the duality horizon). 

All of these observations can be verified by explicit calculations of the tree-level scattering amplitudes \cite{deRham:2013hsa}. 
What about at the quantum level? For this we must remember that all of the theories being considered are effective field theories, and thus each of the actions we have written down should be understood as being supplemented by an infinite number of irrelevant higher derivative interactions. Since all tree-level scattering amplitudes are the same, those parts of the one-loop diagrams which can be determined by unitarity from the tree-level are automatically the same. What is missing are then the contact terms that receive contributions from the higher order EFT interactions. Here, the locality of the dual formulations is what guarantees success. Since the dual of a local theory is itself local, for every counterterm needed in the original theory, there is a dual local counterterm. It thus automatically follows that the loop amplitudes computed perturbatively in the EFT sense can be made to match provided that we allow ourselves the freedom to add independent counterterms on each side of the duality map. 

The subtle question then is, do the coefficients of the EFT counterterms map under the duality transformation. For this, we need to understand whether the regularisation procedure is invariant under duality. Here again it helps us to view things from the perspective of gravitational theory. The core issue is whether the regularisation scheme preserves diffeomorphism invariance. If it does, then the counterterms will naturally and transparently transform under duality. If it does not, then it is an unsuitable choice of regulator for a gravitational theory (or for a theory derived from a gravitational one, or one that is ultimately meant to couple to gravity).

\subsection{Path Integral Measure}

\label{pathintegralmeasure}

To make the equivalence at the quantum level more explicit, we can address the question of how the path integral transforms under the duality. There are two central questions:
\begin{itemize}
\item Is there a regularisation scheme consistent with the duality?
\item Does the path integral measure transform, and can all its effect be accounted for by local counterterms?
\end{itemize}
If the answer to these questions is positive, then we can safely state that as effective field theories the two dual formulations are equivalent. To address this, it is helpful to rewrite the duality entirely in the language of gauge theories, where the answer to these questions are known.

\subsubsection{BRST formulation}

Consider the non-gravitational setting, with a scalar field theory in Minkowski spacetime, the leading form of which is \eqref{general1}. As it stands, the action is not diffeomorphism invariant. To make it so, we may introduce $d$ \stu fields $Y^a(x)$ as in Section~\ref{gaugechoice} which in unitary gauge take the form $Y^a(x)=x^a$. The resulting action is now diffeomorphism invariant and may be quantised via the BRST formalism. Specifically, we introduce Faddeev-Popov-DeWitt ghosts $c^{\mu}$ and $\bar c_{\mu}$ and a Nakanishi-Lautrup field $B_{\mu}$ such that the nilpotent BRST variation is
\ba
&& \hat s Y^a(x) = c^{\mu}(x) \partial_{\mu} Y^a(x) \, , \quad  \hat s c^{\mu}(x) = c^{\nu}(x) \partial_{\nu} c^{\mu}(x)  \, , \\
&& \hat s \phi(x) = c^{\mu}(x) \partial_{\mu} \phi(x) \, ,  \quad \hat s \bar c_{\mu}(x) = B_{\mu}(x)  \, , \quad  \hat s  B_{\mu}(x) = 0 \, .
\ea
In the \stu form, $\mathcal X$ is now defined to be $\mathcal X=-\frac{1}{2} \Phi^a\Phi_a$ so that it too transforms as a scalar
\be
\hat s \mathcal X = c^{\mu}\partial_{\mu} \mathcal X \, ,
\ee
with the diffeomorphism scalars $\Phi^a$ defined via
\be
\Phi_a \d Y^a = \d \phi \, ,
\ee
such that $\hat s \Phi^a(x) = c^{\mu}(x) \partial_{\mu} \Phi^a(x)$.
The BRST path integral is defined via
\be
Z = \int {\cal D}[\phi]\int {\cal D}[Y^a]\int {\cal D}[c^{\mu}]\int {\cal D}[\bar c_{\mu}]\int {\cal D}[B_{\mu}] e^{i S[\phi,Y]+ i \hat s {\cal G}[\phi,Y,c,\bar c,B]} \, ,
\ee
where $S[\phi,Y]$ is diffeomorphism invariant and ${\cal G}$ is the gauge-fixing fermion. If we make the choice
\be
{\cal G} =  \int \d^d x , \bar c_{\mu}(x) (Y^{\mu}(x)-x^{\mu}) \, ,
\ee
then
\be
\hat s {\cal G} = \int \d^d x  [ B_{\mu}(x) (Y^{\mu}(x)-x^{\mu}) - \bar c_{\mu}(x) c^{\nu}(x) \partial_{\nu} Y^{\mu}(x)]\, .
\ee
Now since $S[\phi,Y]-S[\phi,x]$ vanishes when $Y^a=x^a$, then Taylor expanding around this solution, every term is at least linear in $Y^a-x^a$. We can therefore always reorganise the Taylor expansion into the form
\be
S[\phi,Y]-S[\phi,x] = \int \d^d x , J_{\mu}(\phi,Y,x)(Y^{\mu}(x)-x^{\mu})\, .
\ee
If we now perform a field-dependent translation of the Nakanishi-Lautrup field
\be
B_{\mu}(x) = \bar B_{\mu}(x)-\partial_{\nu} (\bar c_{\mu}(x) c^{\nu}(x))-J_{\mu}(\phi,Y,x) \, ,
\ee
then the path integral becomes
\be
Z = \int {\cal D}[\phi]\int {\cal D}[Y^a]\int {\cal D}[c^{\mu}]\int {\cal D}[\bar c_{\mu}]\int {\cal D}[\bar B_{\mu}] e^{i S[\phi,x]+ i  \int \d^d x  [ \bar B_{\mu}(x) (Y^{\mu}(x)-x^{\mu}) - \bar c_{\mu}(x) c^{\mu}(x) ]}\, .
\ee
There is no induced measure because the field redefinition is linear in $B$.
The integral over the Nakanishi-Lautrup field gives a functional delta function which removes the $Y^a$ path integral. The remaining Faddeev-Popov-DeWitt path integral is local and gives a trivial constant, so that we recover the original scalar field theory
\be
Z = \int {\cal D}[\phi] e^{i S[\phi,x]} \, ,
\ee
with no additional measure.

\subsubsection{Duality as a gauge-fixing}

In the BRST formalism, the duality corresponds to the ability to make a different choice of gauge-fixing fermion. To be precise, we now consider
\be
{\cal G} =  \int \d^d x , \bar c_{\mu}(x) (Y^{\mu}(x)-x^{\mu}-\tilde G(\tilde \phi,\tilde{\mathcal X}) \tilde \Phi^{\mu}(x) )\, .
\ee
In addition we are choosing to perform the local invertible field redefinition
\be \label{FR1}
\phi = \tilde \phi + \tilde F[\tilde \phi, \tilde{\mathcal X}]\, .
\ee
Although not necessary, to be consistent with the earlier notation it is helpful to switch the dummy integration label $x$ to $\tilde x$ so that the two `gauges' are clearly related in the sense
\be
Y^a= x^a = \tilde x^a + \tilde G(\tilde \phi,\tilde{\mathcal X}) \tilde \Phi^a(\tilde x) \, ,
\ee
as required. Remembering that $\tilde \Phi^a$ and $\tilde{\mathcal X}$ are defined via
\be
\tilde \Phi^a = \Omega(\phi,\mathcal X) \Phi^a \, , \quad \tilde{\mathcal X} = -\frac{1}{2} \tilde \Phi^a \tilde \Phi_a \, ,
\ee
they are both diffeomorphism scalars. Thus the combination $\tilde G(\tilde \phi,\tilde{\mathcal X}) \tilde \Phi^a(\tilde x)$ is also a diffeomorphism scalar meaning under a BRST transformation
\be
\hat s ( \tilde G(\tilde \phi,\tilde{\mathcal X}) \tilde \Phi^a(\tilde x)) = c^{\mu} \partial_{\mu} (\tilde G(\tilde \phi,\tilde{\mathcal X}) \tilde \Phi^a(\tilde x) )\, .
\ee
The BRST exact part of the action is now
\ba
\hat s {\cal G} &=& \int \d^d \tilde x  [ B_{\mu}(\tilde x) (Y^{\mu}(\tilde x)-\tilde x^{\mu}-\tilde G(\tilde \phi,\tilde{\mathcal X}) \tilde \Phi^\mu(\tilde x))-\bar c_{\mu}(\tilde x) c^{\nu}(\tilde x) \tilde \partial_{\nu} Y^{\mu}(\tilde x)   \\
&+& \bar c_{\mu}(\tilde x) c^{\nu}(\tilde x) \tilde \partial_{\nu }( \tilde G(\tilde \phi,\tilde{\mathcal X}) \tilde \Phi^\mu(\tilde x))
] \, . \nn
\ea

We may perform a field redefinition of the Nakanishi-Lautrup field to remove all the terms that vanish when $Y^{\mu}(\tilde x)-\tilde x^{\mu}-\tilde G(\tilde \phi,\tilde{\mathcal X}) \tilde \Phi^\mu(\tilde x)=0$, with the result that the path integral is
\ba
Z &=& \int {\cal D}[ \phi]\int {\cal D}[Y^a]
\int {\cal D}[c^\mu]\int {\cal D}[\bar c_\mu]
\int {\cal D}[\bar B_\mu] \,
e^{iS[\phi,\tilde x+\tilde G\tilde \Phi]}
\nn \\
&&\times
\exp\left(
i\int d^d\tilde x ,
\left[
\bar B_\mu
\left(
Y^\mu-\tilde x^\mu-\tilde G(\tilde \phi,\tilde{\mathcal X}) \tilde \Phi^{\mu}\right)
-\bar c_\mu c^\mu
\right]
\right)\, .
\ea
The ghost path integral remains trivial in this gauge. The integral over the Nakanishi-Lautrup field gives a functional delta function
\be
\delta(Y^\mu-\tilde x^\mu-\tilde G(\tilde \phi,\tilde{\mathcal X}) \tilde \Phi^{\mu}) \, ,
\ee
but we must remember that $\tilde \Phi^{\mu}$ depends non-trivially on $Y^a$.
It is useful to introduce the dual \stu fields
\be
\tilde Y^a
\equiv
Y^a-\tilde G(\tilde\phi,\tilde{\mathcal X})\tilde\Phi^a\, .
\ee
The defining relation for the dual \stu field is
\be
\tilde\Phi_a \d \tilde Y^a =\tilde\Phi_a \d
\left(Y^a-\tilde G(\tilde\phi,\tilde{\mathcal X})\tilde\Phi^a
\right)=
\d\tilde\phi\, .
\ee
The consistency of this defining relation requires the integrability condition \eqref{eq:condition1} to be satisfied.
The gauge condition is therefore simply $\tilde Y^a(\tilde x)=\tilde x^a $.
On the support of the functional delta function, the defining relation becomes
\be
\tilde\Phi_a \d\tilde x^a=\d\tilde\phi \, ,
\ee
and hence
\be
\tilde\Phi^a=\tilde\partial^a\tilde\phi \, ,
\qquad
\tilde{\mathcal X}
=-\frac{1}{2}
\tilde\partial_\mu\tilde\phi
\tilde\partial^\mu\tilde\phi\, .
\ee
However, the functional delta function should not be used to eliminate the integral of the $Y^a$ path without accounting for its Jacobian. The appropriate procedure is to perform the combined change of variables
\be
(\phi,Y^a)
\longrightarrow
(\tilde\phi,\tilde Y^a)\, .
\ee
The corresponding functional measure is
\be
{\cal D}[\phi] \,{\cal D}[Y^a]
=
{\cal D}[\tilde\phi] \,{\cal D}[\tilde Y^a] \,{\cal J}[\tilde\phi,\tilde Y] \, ,
\ee
where
\be \label{Jacobian1}
{\cal J}[\tilde\phi,\tilde Y]
=
{\rm Det}
\left[
\begin{array}{cc}
\displaystyle
\frac{\delta\phi}{\delta\tilde\phi}
&
\displaystyle
\frac{\delta\phi}{\delta\tilde Y^b} \\
[3mm]
\displaystyle
\frac{\delta Y^a}{\delta\tilde\phi}
&
\displaystyle
\frac{\delta Y^a}{\delta\tilde Y^b}
\end{array}
\right]\, .
\ee
The integral over the Nakanishi-Lautrup field now gives
\be
\delta[\tilde Y^a-\tilde x^a] \, ,
\ee
so that the $\tilde Y^a$ path integral sets $\tilde Y^a=\tilde x^a $.
On this gauge slice,
\be
S[\phi,Y]=S\left[
\tilde\phi+\tilde F(\tilde\phi,\tilde{\mathcal X}),
\tilde x+\tilde G(\tilde\phi,\tilde{\mathcal X})
\tilde\partial\tilde\phi
\right]\equiv
\tilde S[\tilde\phi,\tilde x] \, ,
\ee
which is nothing other than the dual action. The path integral consequently takes the form
\be
Z=
\int {\cal D}[\tilde\phi] ,
\mu[\tilde\phi] ,
e^{i\tilde S[\tilde\phi,\tilde x]} \, ,
\ee
where the complete induced measure is
\be
\mu[\tilde\phi]
=\left.
{\cal J}[\tilde\phi,\tilde Y]
\right|_{\tilde Y=\tilde x}\, .
\ee
The functional Jacobian is the determinant of a matrix of local differential operators. For example, the scalar-scalar block is
\be
\left.
\frac{\delta\phi(\tilde x)}
{\delta\tilde\phi(\tilde x')}
\right|_{\tilde Y=\tilde x}
=
\left[
1+{\cal B}[\tilde\phi]
\right]
\delta^d(\tilde x-\tilde x') \, ,
\ee
where
\be
{\cal B}\delta^d(\tilde x-\tilde x')
=
\frac{\partial\tilde F}
{\partial\tilde\phi(\tilde x)}
\delta^d(\tilde x-\tilde x')
-
\left(
\tilde\partial^\mu\tilde\phi
\frac{\partial\tilde F}{\partial\tilde{\mathcal X}}
\right)
\tilde\partial_\mu
\delta^d(\tilde x-\tilde x')\, .
\ee
The other entries of the full Jacobian are likewise local differential operators. To see this explicitly, define the local vielbein and its inverse
\be
\tilde E_\mu{}^a
\equiv
\tilde\partial_\mu\tilde Y^a,
\qquad
\tilde E_a{}^\mu\tilde E_\mu{}^b
=\delta_a{}^b,
\qquad
\tilde E_\mu{}^a\tilde E_a{}^\nu
=\delta_\mu{}^\nu\, .
\ee
Then the dual field $\tilde \Phi^a$ is
\be
\tilde\Phi_a
=
\tilde E_a{}^\mu\tilde\partial_\mu\tilde\phi\, .
\ee
At fixed $\tilde\phi$, its variation with respect to $\tilde Y$ is
\be
\delta\tilde\Phi_a=
-\tilde E_a{}^\mu\tilde\Phi_b
\tilde\partial_\mu\delta\tilde Y^b\, .
\ee
In particular, on the gauge slice $\tilde Y^a=\tilde x^a$,
\be
\delta\tilde\Phi^a =
-\tilde\Phi_b
\tilde\partial^a\delta\tilde Y^b \, , \quad
\delta\tilde{\mathcal X}
=
\tilde\Phi^\mu\tilde\Phi_b
\tilde\partial_\mu\delta\tilde Y^b\, .
\ee
Thus, all entries of the full Jacobian contain only local functions with a finite number of derivatives acting on delta functions.

Perturbatively, the logarithm of the complete determinant may be expanded as
\be
\ln\mu
= {\rm Tr}\ln[1+{\mathbb B}]=
{\rm Tr}[{\mathbb B}]
-\frac{1}{2}{\rm Tr}[{\mathbb B}^2]
+\frac{1}{3}{\rm Tr}[{\mathbb B}^3]
+\dots \, ,
\ee
where ${\mathbb B}$ denotes the matrix-valued local differential operator obtained from the full Jacobian. Each functional trace reduces to a sum of local background field expressions multiplying momentum integrals of the form
\be
\int\frac{\d^d k}{(2\pi)^d}
k^{\mu_1}\dots k^{\mu_n}\, .
\ee
There are no propagator denominators in these integrals. They are therefore pure power-law divergences and vanish in dimensional regularisation,
\be
\int\frac{\d^d k}{(2\pi)^d}
k^{\mu_1}\dots k^{\mu_n}=0\, .
\ee
For example, focusing on the scalar block
\ba
{\rm Tr}[{\cal B}]
&=&
\int \d^d\tilde x
\int\frac{\d^d k}{(2\pi)^d}
\left[
\frac{\partial\tilde F}
{\partial\tilde\phi(\tilde x)}
+
i k_\mu
\left(
\tilde\partial^\mu\tilde\phi
\frac{\partial\tilde F}{\partial\tilde{\mathcal X}}
\right)
\right]
\nn\\
&=&0\, .
\ea
The term linear in $k_\mu$ vanishes by symmetric integration, while the remaining term is a scaleless power-divergent integral. The same type of argument applies at every order in the perturbative expansion.
Consequently, up to an irrelevant field-independent normalisation, in dimensional regularisation
\be
\mu[\tilde\phi]\equiv1\, .
\ee
We therefore obtain
\be
Z=\int {\cal D}[\phi] e^{iS[\phi,x]}
=\int {\cal D}[\tilde\phi]
e^{i\tilde S[\tilde\phi,\tilde x]}\, .
\ee
Together with the invertibility of the transformation and the usual
assumptions of the equivalence theorem, this establishes the equivalence
of the $S$-matrix. Had we not chosen dimensional regularisation, then we would find that $\ln \mu[\tilde\phi]$ would be determined by the pure power-law divergent terms. But each of these terms is manifestly local, meaning it can be removed with a local counterterm. Thus in matching the EFT and its dual, we simply need to allow for the appropriate renormalisation of the local higher dimension operators. 

As we have seen, the $S$-matrix is equivalent. Local off-shell correlation functions, on the other hand, are not diffeomorphism invariant, and so do transform non-trivially under the duality. For example, the generating functional that includes sources for $\phi$ transforms as
\ba
Z[J] &=& \int {\cal D}[ \phi] e^{i S[ \phi, x]+i \int \d^d x J(x)  \phi( x)} \nn \\
&=&\int {\cal D}[\tilde \phi] e^{i \tilde S[\tilde \phi,\tilde x]+i \int \d^d \tilde x {\rm det}[\eta^{\mu}_{\nu}+\tilde \partial_{\nu} (\tilde G \tilde \partial^{\mu} \tilde \phi)]J(\tilde x+\tilde G(\tilde \phi,\tilde{\mathcal X}) \tilde \partial \tilde \phi) (\tilde \phi(\tilde x)+\tilde F(\tilde \phi,\tilde{\mathcal X}))}\, .
\ea
The field-dependence of the argument of the external source implies that local off-shell correlation functions in one formalism map to non-local off-shell correlation functions in the other \cite{Keltner:2015xda}. This phenomenon is a characteristic of a gravitational theory where there are no local gauge invariant correlation functions. It is precisely because of this that modern discussions highlight the importance of analyticity of scattering amplitudes in establishing causality and locality.

\subsubsection{Path Integral Measure with Gravity}

Returning to the case where gravity is dynamical (and massless), we no longer need to introduce \stu fields $Y^a(x)$. We can continue to use the BRST formalism, now extended to include the local Lorentz symmetry. Rather than a choice on $Y^a(x)$, the natural gauge fixings would then amount to gauge choices on the vielbein $e^a(x)$ and scalar $\phi(x)$ and potentially other fields. Beyond the usual BRST factors, the only thing we need to keep track of is the purely local field redefinition of the vielbein and scalar field
\ba
&& \phi=\tilde\phi+\tilde F(\tilde\phi,\tilde{\mathcal X})\, ,
\nn\\
&& e^a =
\tilde e^a
+D[\tilde\omega](\tilde G(\tilde\phi,\tilde{\mathcal X})
\tilde\Phi^a)\, ,
\nn\\
&& \omega^{ab} = \tilde \omega^{ab}\, .
\ea
Put differently, the corresponding path integral measure arises from
\be
{\cal D}[\phi] \,{\cal D}[e^a] {\cal D}[\omega^{ab}]
=
{\cal D}[\tilde \phi] \,{\cal D}[\tilde e^a] {\cal D}[\tilde \omega^{ab}] {\cal J}[\tilde\phi,\tilde e , \tilde  \omega]\, ,
\ee
where
\be
{\cal J}[\tilde\phi,\tilde e , \tilde  \omega]
=
{\rm Det}
\left[
\begin{array}{ccc}
\displaystyle
\frac{\delta\phi}{\delta\tilde\phi}
&
\displaystyle
\frac{\delta\phi}{\delta\tilde e^b}
&
\displaystyle
0 \\
[3mm]
\displaystyle
\frac{\delta e^a}{\delta\tilde\phi}
&
\displaystyle
\frac{\delta e^a}{\delta\tilde e^b}
&
\displaystyle
\frac{\delta e^a}{\delta\tilde\omega^{bc}} \\
[3mm]
0
&
0
&
\displaystyle
\frac{\delta\omega^{ab}}{\delta\tilde\omega^{cd}}
\end{array}
\right] .
\ee
Since the spin connection is unchanged, the last block is just the identity,
\be
\frac{\delta\omega^{ab}(x)}
{\delta\tilde\omega^{cd}(x')}
=
\delta^{ab}_{cd} \, \delta^d(x-x')\, .
\ee
Thus, the only possible field-dependent contribution comes from the
$(\phi,e^a)$ blocks. This is exactly what corresponds in the decoupling limit to the Jacobian \eqref{Jacobian1}.
To make the locality of this determinant explicit, introduce the inverse vielbein
\be
\tilde E_a{}^\mu \tilde e^b{}_\mu=\delta_a{}^b,
\qquad
\tilde E_a{}^\mu \tilde e^a{}_\nu=\delta^\mu{}_\nu ,
\ee
so that
\be
\tilde\Phi_a=\tilde E_a{}^\mu\partial_\mu\tilde\phi,
\qquad
\tilde{\mathcal X}=-\frac12\tilde\Phi_a\tilde\Phi^a .
\ee
The variations of the auxiliary scalar variables are then
\ba
\delta\tilde\Phi_a
&=&
\tilde E_a{}^\mu\partial_\mu\delta\tilde\phi
-
\tilde E_a{}^\mu\tilde\Phi_b\,
\delta\tilde e^b{}_\mu ,
\nn\\
\delta\tilde{\mathcal X}
&=&
-\tilde E_a{}^\mu\tilde\Phi^a\partial_\mu\delta\tilde\phi
+
\tilde E_a{}^\mu\tilde\Phi^a\tilde\Phi_b\,
\delta\tilde e^b{}_\mu .
\ea
It follows that the scalar part of the field redefinition varies as
\ba
\delta\phi
&=&
\left(
1+\tilde F_{,\tilde\phi}
-
\tilde F_{,\tilde{\mathcal X}}\tilde E_a{}^\mu\tilde\Phi^a\partial_\mu
\right)
\delta\tilde\phi
+
\tilde F_{,\tilde{\mathcal X}}\tilde E_a{}^\mu\tilde\Phi^a\tilde\Phi_b
\delta\tilde e^b{}_\mu .
\ea
For the vielbein block, it is useful to first vary the vector
$\tilde G\tilde\Phi^a$. One finds
\ba
\delta(\tilde G\tilde\Phi^a)
&=&
{\cal C}^a{}_\phi\,\delta\tilde\phi
+
{\cal C}^{a\mu}{}_{b}\,\delta\tilde e^b{}_\mu ,
\ea
where
\ba
{\cal C}^a{}_\phi
&=&
\tilde G_{,\tilde\phi}\tilde\Phi^a
+
\left(
\tilde G\,\tilde E^{a\mu}
-
\tilde G_{,\tilde{\mathcal X}}\tilde\Phi^a\tilde E_b{}^\mu\tilde\Phi^b
\right)
\partial_\mu ,
\nn\\
{\cal C}^{a\mu}{}_{b}
&=&
\tilde G_{,\tilde{\mathcal X}}\tilde\Phi^a\tilde E_c{}^\mu\tilde\Phi^c\tilde\Phi_b
-
\tilde G\,\tilde E^{a\mu}\tilde\Phi_b .
\ea
Therefore
\ba
\delta e^a
&=&
\delta\tilde e^a
+
D[\tilde\omega]
\left(
{\cal C}^a{}_\phi\,\delta\tilde\phi
+
{\cal C}^{a\mu}{}_{b}\,\delta\tilde e^b{}_\mu
\right)
+
\delta\tilde\omega^a{}_{b}\,
\tilde G\tilde\Phi^b .
\ea
We can see by inspection that all the operators that will enter the Jacobian are local with a finite number of derivatives.
Thus, the locality of the gravitational duality ensures that there is no
non-trivial, non-local measure factor, \ie
\be
{\cal J}[\tilde\phi,\tilde e,\tilde\omega]\equiv 1
\ee
in dimensional regularisation, up to local counterterms. The gravitational
duality therefore maps the path integral into the path integral of the dual
theory without producing any additional physical measure contribution.

\subsection{Duality (Gribov) Horizon}

As noted previously, the duality transformation relies on the invertibility of the field-dependent coordinate transformation. In general, one should not expect this transformation to be globally invertible over the entire field space. In the BRST framework, this invertibility issue manifests itself as the question of whether the chosen gauge-fixing fermion can be defined globally. When the transformation fails to be invertible, the associated Faddeev–Popov operator develops a zero mode, and its determinant vanishes. The gauge slice then ceases to define a unique local representative.
This situation is directly analogous to the well-known Gribov problem encountered in ordinary Yang–Mills gauge theories.

This does not constitute an obstruction to the duality within the regime of validity of the effective field theory, since all of the functions other than the purely local $\phi$ dependent field redefinition part, are higher derivative irrelevant operators, and so must be treated perturbatively.
The duality need not furnish a single global coordinate system on the full space of field configurations. Rather, it is sufficient that the transformation define an invertible local chart on the region of field space containing the background and asymptotic states about which the effective theory is perturbatively defined. 
The appearance of a duality horizon therefore signals the boundary of a particular choice of field-space coordinates, rather than a failure of the underlying physical equivalence. Beyond this boundary, one may need to pass to a different gauge choice, or equivalently to a different local parameterisation of field space, just as multiple coordinate charts are required to cover a manifold. 

Although we shall not consider this here, it is possible that the duality could be extended beyond the duality horizon by using methods closely analogous to those used in gauge theories, for example the Gribov-Zwanziger approach \cite{Gribov:1977wm,Zwanziger:1989mf,Vandersickel:2012tz}. It is also plausible that, as in gauge theories, the horizon may disappear with a more suitable choice of gauge fixing terms that do not arise from the Faddeev-Popov argument but are nevertheless consistent within the BRST formalism \cite{Scholtz:1997jp,Rogers:1999zj}.

\section{Discussion}

We have shown that every scalar field theory in Minkowski spacetime admits a duality transformation to another superficially distinct scalar theory that is nevertheless physically equivalent, in the sense that it has an identical S-matrix. The duality is not just a simple field redefinition; it also involves a field-dependent diffeomorphism, which makes the transformation fundamentally nonlocal. However, it remains perturbatively local and does not conflict with the assumptions underlying the LSZ theorem meaning that scattering amplitudes are left invariant. A special case of this diffeomorphic scalar duality was first observed in the class of Galileon theories \cite{Curtright:2012gx,BiHiguchi,deRham:2013hsa,deRham:2014lqa}, and later in DBI-Galileon theories \cite{Chagoya:2016jyn}. The existence of a duality transformation requires only the solution of an integrability condition, identified in Section~\ref{Integrability}, and the invertibility of the  associated with the diffeomorphism. Explicit global solutions of the former are given in Sections~\ref{Example1} and \ref{Example2}.
The invertibility of the Jacobian is violated at what we refer to as the duality horizon. Generically, the duality horizon lies outside the regime of validity of the effective field theory and is therefore not a concern; see Section~\ref{dualityhorizon}. The scalar may easily be coupled to matter of arbitrary spin without spoiling the duality properties.

It was previously known that a special case of the duality emerges from massive gravity theories \cite{BiHiguchi,deRham:2014lqa}. In the present work, we have shown how to couple the scalar field theories to massless gravity in a way that makes the duality manifest. We find a very general class of first-order gravitational theories that map into themselves under the duality transformation.

As in the massive gravity context, the duality emerges from a diffeomorphism in the gravitational theory that relates two superficially different decoupling limits, or weak-field gravity limits. On the gravitational side, the equivalence is manifest, since the two descriptions are related by a combination of a local field redefinition and a gauge transformation, both of which preserve the S-matrix. In the decoupling limit, the duality appears as a nonlocal field-dependent diffeomorphism. A summary of the duality transformations in both the non-gravitational and gravitational contexts is given in Table~\ref{tab:dualitysummary}.

At no stage do we assign any role to the possible existence of nonlinearly realised symmetries in the decoupling limit, thereby obviating the need for the coset construction. However, as noted in Section~\ref{Example2}, special duality transformations of the more general class presented here can exhibit symmetry group transformation properties.

In general, the local field redefinition of the vielbein breaks the torsion-free condition of Einstein-Hilbert gravity. Consequently, the dual of a theory with matter minimally coupled to Einstein gravity corresponds to a theory with torsion. This appears in the decoupling limit through the existence of non-standard kinetic mixings between the scalar and the massless spin-2 graviton, some of which cannot be demixed by a local field redefinition.\footnote{This phenomenon is well known in the context of massive gravity theories \cite{deRham:2014zqa}, where one of the terms in the decoupling limit of dRGT massive gravity \cite{deRham:2010kj} mixes the helicity-2 and helicity-0 modes in a way that cannot be removed by a local field redefinition.} Far from being a problem, such couplings are natural to consider within an extended gravitational EFT.
It would be very interesting to explore the phenomenological implications of such couplings in the context of gravitational waves propagating in dark energy theories, where they may naturally lead to induced scalar gravitational waves. This duality may also have deeper formal implications in generic field theories, causality considerations and generalised symmetries which would be worth exploring. 
We leave such considerations to future work.

\begin{table}[H]
\centering
\renewcommand{\arraystretch}{1.20}
\begin{tabular}{|c|c|c|}
\hline
& \textbf{Minkowski/decoupling limit} & \textbf{Gravitational theory} \\
\hline\hline
Variables
&
\begin{minipage}[c]{0.35\textwidth}
\ba
&& x^\mu,\quad \phi,\quad X=-\frac12 \p_\mu\phi\p^\mu\phi . \nn
\ea
\end{minipage}
&
\begin{minipage}[c]{0.35\textwidth}
\ba
&& e^a,\quad \omega^{ab},\quad \phi,\quad \Phi^a, \nn \\
&& e^a\Phi_a=\d\phi,\quad
\mathcal X=-\frac12\Phi_a\Phi^a . \nn \\ \nn
\ea
\end{minipage}
\\
\hline
Forward map
&
\begin{minipage}[c]{0.35\textwidth}\vspace{-0.2cm}
\ba
&& \tilde\phi
=\phi+F,
\nn\\
&& \tilde\p_\mu\tilde\phi = \Omega\, \p_\mu\phi ,
\quad \tilde X=\Omega^2X , \nn \\[0.3cm]
&& \tilde x^\mu = x^\mu+G\, \p^\mu\phi .
\nn
\\ \nn
\ea
\end{minipage}
&
\begin{minipage}[c]{0.35\textwidth}
\ba
&& \tilde\phi
=
\phi+F,
\nn\\
&& \tilde\Phi^a
=\Omega\, \Phi^a \, , \quad  \tilde{\mathcal X}
=\Omega^2\mathcal X,
\nn\\
&& \tilde e^a
=e^a+D[\omega](G\Phi^a),
\nn\\
&& \tilde\omega^{ab}=\omega^{ab}. \nn \\ \nn
\ea
\end{minipage}
\\
\hline
Inverse map
&
\begin{minipage}[c]{0.35\textwidth}\vspace{-0.2cm}
\ba
&& \phi = \tilde\phi+\tilde F,
\nn\\
&& \p_\mu\phi =
\tilde\Omega\, 
\tilde\p_\mu\tilde\phi,
\quad  X=\tilde\Omega^2\tilde X , \nn \\[0.3cm] 
&& x^\mu=\tilde x^\mu
+\tilde G\, \tilde\p^\mu\tilde\phi.
\nn\\
\nn
\ea
\end{minipage}
&
\begin{minipage}[c]{0.35\textwidth}
\ba
&& \phi=\tilde\phi+\tilde F,
\nn\\
&& \Phi^a
=
\tilde\Omega\tilde\Phi^a \, , \quad \mathcal X
=
\tilde\Omega^2\tilde{\mathcal X},
\nn\\
&& e^a
=
\tilde e^a
+D[\tilde\omega](\tilde G
\tilde\Phi^a),
\nn\\
&& \omega^{ab}=\tilde\omega^{ab}. \nn \\ \nn
\ea
\end{minipage}
\\
\hline
Inverse functions
&
\begin{minipage}[c]{0.35\textwidth}
\ba
(\tilde F,\tilde G,\tilde\Omega)=(-F,-G\Omega^{-1},\Omega^{-1}) \nn 
\ea\vspace{-0.2cm}
\end{minipage}
&
\begin{minipage}[c]{0.35\textwidth}
\ba
(\tilde F,\tilde G,\tilde\Omega)=(-F,-G\Omega^{-1},\Omega^{-1}) \nn 
\ea\vspace{-0.2cm}
\end{minipage}
\\
\hline
Integrability
&
\begin{minipage}[c]{0.35\textwidth}
\be
\d\phi+\d F
=\Omega\d\phi
-2\Omega X\d G
-\Omega G\d X  \nn 
\ee 
\vspace{-0.2cm}
\end{minipage}
&
\begin{minipage}[c]{0.35\textwidth}
\be
\d\phi+\d F
=\Omega\d\phi
-2\Omega\mathcal X\d G
-\Omega G\d\mathcal X  \nn 
\ee 
\vspace{-0.2cm}
\end{minipage}
\\
\hline
Invertibility
&
\begin{minipage}[c]{0.35\textwidth}
\be
\det\left[
\delta^\mu{}_\nu
+
\p_\nu(G\p^\mu\phi)
\right]\neq0 . \nn \\ \nn 
\ee 
\be
\det\left[
\delta^\mu{}_\nu
+
\tilde \p_\nu(\tilde G \tilde \p^\mu \tilde \phi)
\right]\neq0 . \nn \\ \nn 
\ee 
\vspace{0.5pt}
\end{minipage} 
&
\begin{minipage}[c]{0.37\textwidth}\vspace{-0.8cm}
\centering
\be 
(e^a,\omega^{ab},\phi,\Phi^a)
\longleftrightarrow
(\tilde e^a,\tilde\omega^{ab},\tilde\phi,\tilde\Phi^a) \nn \vspace{-0.1cm}
\ee 
Invertible local field redefinition. \\
\end{minipage}
\\
\hline
Spin-2 fields 
&
\begin{minipage}[c]{0.35\textwidth}
\ba
&& h^a=
\tilde h^a
+\tilde\mu^a{}_b
\tilde G
\tilde\p^b\tilde\phi,
\nn\\
&& \tilde h^a=
h^a
+\mu^a{}_bG\p^b\phi,
\nn\\
&& \tilde\mu^{ab}=\mu^{ab}. \nn \\ \nn
\ea 
\end{minipage} 
&
\begin{minipage}[c]{0.35\textwidth}
\ba
&& h^a=
\tilde h^a
+\tilde\mu^a{}_b
\tilde G
\tilde \Phi^b,
\nn\\
&& \tilde h^a=
h^a
+\mu^a{}_bG\Phi^b,
\nn\\
&& \tilde\mu^{ab}=\mu^{ab}. \nn \\ \nn
\ea
\end{minipage}
\\
\hline
\end{tabular}
\caption{Summary of the diffeomorphic scalar duality in the Minkowski/decoupling-limit formulation and in the first-order gravitational formulation. In the gravitational theory the field-dependent diffeomorphism is replaced by an invertible local field redefinition of the vielbein, spin-connection and scalar field. The functions $F,G$ and $
\Omega$ depend on $\phi,X$ in Minkowski and $\phi,{\cal X}$ in the gravitational theory.
Similarly $\tilde F,\tilde G,\tilde \Omega$ depend on $\tilde \phi,\tilde X$ or $\tilde \phi,\tilde {\cal X}$ respectively.}
\label{tab:dualitysummary}
\end{table}

\section*{Acknowledgments}

The work of AJT and CdR is supported by STFC Consolidated Grant ST/X000575/1. CdR is also supported by Simons Investigator award 690508. 
AJT and CdR would like to thank Ratatouille for stimulating insights.

\appendix

\section{Closure of the Symmetry Transformation}

\label{app:closure}

We now verify that the specific example of the duality considered in Section~\ref{Example2} forms a symmetry group. For the family
\be
\tilde X=f_\alpha(X)
=\sinh\left(e^\alpha\operatorname{arcsinh}X\right)\, ,
\qquad
\Omega_\alpha^2(X)=\frac{f_\alpha(X)}{X}\, ,
\ee
we have the composition property
\be
f_{\alpha+\beta}(X)=f_\beta(f_\alpha(X))\,.
\ee
It follows immediately that
\be
\Omega_{\alpha+\beta}(X)=\Omega_\alpha(X)\Omega_\beta(\tilde X)\, ,
\ee
where we have chosen the same branch of the square root throughout. It is useful to define
\be
D_\alpha(X)
\equiv
\Omega_\alpha(X)+2X\Omega_{\alpha,X}(X)\,.
\ee
Since
\be
\frac{\d \tilde X}{\d X}=\Omega_\alpha D_\alpha \,,
\ee
the composition law for $f_\alpha$ also implies
\be
D_{\alpha+\beta}(X)=D_\alpha(X)D_\beta(\tilde X)\,.
\ee
For the choice $G_0=F_c=0$, the scalar transformation is
\be
\tilde\phi
=\phi \, 
\frac{\Omega_\alpha^2(X)}{D_\alpha(X)}\,,
\ee
and the coordinate shift is
\be
\tilde x^\mu=x^\mu+G_\alpha(\phi,X)\p^\mu\phi \,,
\qquad
G_\alpha(\phi,X)=\phi \, \frac{\Omega_{\alpha,X}(X)}{D_\alpha(X)} \,.
\ee
The closure of the symmetry transformation requires
\be
G_{\alpha+\beta}(\phi,X)=G_\alpha(\phi,X)+
\Omega_\alpha(X)G_\beta(\tilde\phi,\tilde X)\,.
\ee
We shall now verify this identity. Differentiating
\be
\Omega_{\alpha+\beta}(X)=\Omega_\alpha(X)\Omega_\beta(\tilde X(X)) \, ,
\ee
with respect to $X$ gives
\ba
\Omega_{\alpha+\beta,X}(X)&=&\Omega_{\alpha,X}(X)\Omega_\beta(\tilde X)
+\Omega_\alpha(X)\Omega_{\beta,\tilde X}(\tilde X)
\frac{\d\tilde X}{\d X}\\
&=&\Omega_{\alpha,X}(X)\Omega_\beta(\tilde X)
+\Omega_\alpha^2(X)D_\alpha(X)\Omega_{\beta,\tilde X}(\tilde X)\,.
\ea
Therefore
\ba
G_{\alpha+\beta}(\phi,X)
&=& \phi\,
\frac{\Omega_{\alpha+\beta,X}(X)}{D_{\alpha+\beta}(X)} \, ,
\nn\\
&=&
\phi\,
\frac{
\Omega_{\alpha,X}(X)\Omega_\beta(\tilde X)
+
\Omega_\alpha^2(X)D_\alpha(X)\Omega_{\beta,\tilde X}(\tilde X)
}{D_\alpha(X)D_\beta(\tilde X)} \, ,
\nn\\
&=&
\phi\,\frac{\Omega_{\alpha,X}(X)}{D_\alpha(X)}
+
\Omega_\alpha(X)
\left(
\phi\,\frac{\Omega_\alpha^2(X)}{D_\alpha(X)}
\frac{\Omega_{\beta,\tilde X}(\tilde X)}{D_\beta(\tilde X)}
\right) \, ,
\nn\\
&=&
\phi\,\frac{\Omega_{\alpha,X}(X)}{D_\alpha(X)}
+
\Omega_\alpha(X)
\left(
\tilde \phi \, 
\frac{\Omega_{\beta,\tilde X}(\tilde X)}{D_\beta(\tilde X)}
\right) \, , \nn \\
&=& G_\alpha(\phi,X)+
\Omega_\alpha(X)G_\beta(\tilde\phi,\tilde X)\,.
\ea
In the third-to-last step we used
\be
D_\beta(\tilde X)-\Omega_\beta(\tilde X)
=
2\tilde X\Omega_{\beta,\tilde X}(\tilde X),
\quad
D_\alpha(X)-\Omega_\alpha(X)
=
2X\Omega_{\alpha,X}(X),
\quad
\tilde X=\Omega_\alpha^2(X)X\,.
\ee
This confirms the desired identity.

\section{Derivative identities}

\label{appendix}

To prove the following derivative identity used in \eqref{eq:deri}, and in the rest of Section~\ref{Galileons},
\ba
&& \partial^{\rho} \left( \partial_{\rho} \pi \epsilon \epsilon \eta^{d-n+1} (\partial \partial \pi )^{n-1} \right)
+  (n-1)  \epsilon \epsilon (\partial \partial X) \eta^{d-n+1} (\partial \partial \pi )^{n-2}
\nn \\
&&= (\Box \pi)\epsilon \epsilon \eta^{d-n+1} (\partial \partial \pi )^{n-1}
+ (n-1) \partial_{\rho} \pi \epsilon \epsilon \eta^{d-n+1} (\partial \partial \pi )^{n-2}
\partial \partial \partial^{\rho} \pi
\nn\\
&& \quad \quad
+ (n-1) \epsilon \epsilon
\left(
-\partial_{\mu} \pi \partial \partial \partial^{\mu} \pi
-\partial \partial_{\mu} \pi \partial \partial^{\mu} \pi
\right)
\eta^{d-n+1} (\partial \partial \pi )^{n-2}
\nn\\
&& = (\Box \pi)\epsilon \epsilon \eta^{d-n+1} (\partial \partial \pi )^{n-1}
-(n-1) \epsilon \epsilon
\partial \partial_{\mu} \pi
\partial \partial^{\mu} \pi
\eta^{d-n+1} (\partial \partial \pi )^{n-2}
\nn \\
&& =(d-n+1) \epsilon \epsilon \eta^{d-n} (\partial \partial \pi )^{n} \, ,
\ea
it is helpful to  first spell out the index conventions. We define $
\Pi_{\mu\nu}\equiv \partial_\mu\partial_\nu\pi$, 
$\Pi^\mu{}_\nu\equiv \eta^{\mu\alpha}\Pi_{\alpha\nu}$, $X=-\frac12\partial_\mu\pi\,\partial^\mu\pi$.
The double-epsilon notation is defined such that for any collection
of mixed tensors $A_i{}^\mu{}_\nu$,
\be
\epsilon\epsilon\,\delta^{d-k} A_1\cdots A_k
\equiv
\epsilon_{\mu_1\cdots\mu_d}
\epsilon^{\nu_1\cdots\nu_d}
\delta^{\mu_1}{}_{\nu_1}\cdots
\delta^{\mu_{d-k}}{}_{\nu_{d-k}}
(A_1)^{\mu_{d-k+1}}{}_{\nu_{d-k+1}}
\cdots
(A_k)^{\mu_d}{}_{\nu_d}.
\ee
In particular,
\be
\epsilon\epsilon\,\delta^{d-n+1}\Pi^{n-1}
\equiv
\epsilon_{\mu_1\cdots\mu_d}
\epsilon^{\nu_1\cdots\nu_d}
\delta^{\mu_1}{}_{\nu_1}\cdots
\delta^{\mu_{d-n+1}}{}_{\nu_{d-n+1}}
\Pi^{\mu_{d-n+2}}{}_{\nu_{d-n+2}}
\cdots
\Pi^{\mu_d}{}_{\nu_d},
\ee
and
\be
\epsilon\epsilon\,\delta^{d-n}\Pi^n
\equiv
\epsilon_{\mu_1\cdots\mu_d}
\epsilon^{\nu_1\cdots\nu_d}
\delta^{\mu_1}{}_{\nu_1}\cdots
\delta^{\mu_{d-n}}{}_{\nu_{d-n}}
\Pi^{\mu_{d-n+1}}{}_{\nu_{d-n+1}}
\cdots
\Pi^{\mu_d}{}_{\nu_d}.
\ee
The term containing two explicit contracted derivatives is therefore
\ba
&&
\epsilon\epsilon\,
\partial\partial_\rho\pi\,
\partial\partial^\rho\pi\,
\delta^{d-n+1}
\Pi^{n-2}
\nn\\
&\equiv&
\epsilon_{\mu_1\cdots\mu_d}
\epsilon^{\nu_1\cdots\nu_d}
\delta^{\mu_1}{}_{\nu_1}\cdots
\delta^{\mu_{d-n+1}}{}_{\nu_{d-n+1}}
\left(
\partial^{\mu_{d-n+2}}\partial_\rho\pi
\right)
\left(
\partial_{\nu_{d-n+2}}\partial^\rho\pi
\right)
\nn\\
&&\times
\Pi^{\mu_{d-n+3}}{}_{\nu_{d-n+3}}
\cdots
\Pi^{\mu_d}{}_{\nu_d}.
\ea
The first equality follows by expanding the total derivative and using
\be
\partial_\alpha\partial_\beta X
=
-\partial_\rho\pi\,\partial_\alpha\partial_\beta\partial^\rho\pi
-\partial_\alpha\partial_\rho\pi\,\partial_\beta\partial^\rho\pi .
\ee
Indeed,
\be
\partial^\rho\Pi_{\mu\nu}
=
\partial_\mu\partial_\nu\partial^\rho\pi ,
\ee
so differentiating one of the $(n-1)$ factors of $\Pi_{\mu\nu}$ gives
$
(n-1)\partial_\rho\pi\,
\epsilon\epsilon\,\delta^{d-n+1}\Pi^{n-2}
\partial\partial\partial^\rho\pi .
$
This cancels the corresponding cubic-derivative term coming from
$\partial\partial X$, since
$
\partial_\rho\pi\,\partial\partial\partial^\rho\pi-\partial_\mu\pi\,\partial\partial\partial^\mu\pi
=0 .
$
Therefore only the quadratic term in $\partial\partial X$ remains, giving
\ba
&&
(\Box \pi)\epsilon\epsilon\,\delta^{d-n+1}\Pi^{n-1}
-(n-1)
\epsilon\epsilon\,\delta^{d-n+1}\Pi^2\Pi^{n-2}.
\ea
It remains to prove the purely algebraic identity
\be
(\Box \pi)\epsilon\epsilon\,\delta^{d-n+1}\Pi^{n-1}
-(n-1)
\epsilon\epsilon\,\delta^{d-n+1}\Pi^2\Pi^{n-2}
=
(d-n+1)\epsilon\epsilon\,\delta^{d-n}\Pi^n .
\ee
In explicit index notation this is
\ba
&&
(\Box\pi)\,
\epsilon_{\mu_1\cdots\mu_d}
\epsilon^{\nu_1\cdots\nu_d}
\delta^{\mu_1}{}_{\nu_1}\cdots
\delta^{\mu_{d-n+1}}{}_{\nu_{d-n+1}}
\Pi^{\mu_{d-n+2}}{}_{\nu_{d-n+2}}
\cdots
\Pi^{\mu_d}{}_{\nu_d}
\nn\\
&&
-(n-1)
\epsilon_{\mu_1\cdots\mu_d}
\epsilon^{\nu_1\cdots\nu_d}
\delta^{\mu_1}{}_{\nu_1}\cdots
\delta^{\mu_{d-n+1}}{}_{\nu_{d-n+1}}
(\Pi^2)^{\mu_{d-n+2}}{}_{\nu_{d-n+2}}
\Pi^{\mu_{d-n+3}}{}_{\nu_{d-n+3}}
\cdots
\Pi^{\mu_d}{}_{\nu_d}
\nn\\
&&=
(d-n+1)
\epsilon_{\mu_1\cdots\mu_d}
\epsilon^{\nu_1\cdots\nu_d}
\delta^{\mu_1}{}_{\nu_1}\cdots
\delta^{\mu_{d-n}}{}_{\nu_{d-n}}
\Pi^{\mu_{d-n+1}}{}_{\nu_{d-n+1}}
\cdots
\Pi^{\mu_d}{}_{\nu_d}.
\ea
The Levi-Civita identities used here are,\footnote{As in the main text we assume one Levi-Civita always has indices down with $\epsilon_{0\dots d-1}=+1$, and the other always up with $\epsilon^{0\dots d-1}=+1$. This avoids the need for minus signs associated with the signature at the price of a slight abuse of notation. Alternatively, one may regard these as the conventions in the Euclidean.}
\be
\epsilon_{\mu_1\cdots\mu_d}
\epsilon^{\nu_1\cdots\nu_d}
=
d!\,
\delta^{[\nu_1}_{\mu_1}\cdots
\delta^{\nu_d]}_{\mu_d},
\ee
and more generally
\be
\epsilon_{\mu_1\cdots\mu_k\alpha_{k+1}\cdots\alpha_d} \epsilon^{\nu_1\cdots\nu_k\alpha_{k+1}\cdots\alpha_d}
=
(d-k)!\,k!\,
\delta^{[\nu_1}_{\mu_1}\cdots
\delta^{\nu_k]}_{\mu_k}.
\ee
The identity above is a standard consequence of the antisymmetry enforced by
the generalised Kronecker delta. Inserting an additional $\Pi$ into any of the
$d-n+1$ metric slots gives the same contribution. Therefore, the right-hand
side carries the overall factor $d-n+1$.

As a consistency check, the coefficient can be fixed by evaluating both sides in the
special configuration $\Pi_{\mu\nu}=\lambda \delta_{\mu\nu}.
$
In this configuration,
$
\Box\pi=d\lambda$, $\Pi^\mu{}_\nu=\lambda\delta^\mu{}_\nu$, $(\Pi^2)^\mu{}_\nu=\lambda^2\delta^\mu{}_\nu$.
Therefore
\ba
\epsilon\epsilon\,\delta^{d-n+1}\Pi^{n-1}
&=&
\lambda^{n-1}\,
\epsilon\epsilon\,\delta^d,
\nn\\
\epsilon\epsilon\,\delta^{d-n+1}\Pi^2\Pi^{n-2}
&=&
\lambda^n\,
\epsilon\epsilon\,\delta^d,
\nn\\
\epsilon\epsilon\,\delta^{d-n}\Pi^n
&=&
\lambda^n\,
\epsilon\epsilon\,\delta^d .
\ea
The left-hand side of the algebraic identity is then
\ba
&&
d\lambda\,
\epsilon\epsilon\,\delta^{d-n+1}\Pi^{n-1}
-(n-1) \epsilon\epsilon\,\delta^{d-n+1}\Pi^2\Pi^{n-2}
\nn\\
&=&
d\lambda^n\,\epsilon\epsilon\,\delta^d-(n-1)\lambda^n\,\epsilon\epsilon\,\delta^d
\nn\\
&=&
(d-n+1)\lambda^n\,\epsilon\epsilon\,\delta^d
\nn\\
&=& (d-n+1)\epsilon\epsilon\,\delta^{d-n}\Pi^n .
\ea
This fixes precisely the coefficient in the final line. Hence
\be
(\Box \pi)\epsilon\epsilon\,\delta^{d-n+1}\Pi^{n-1}
-(n-1)\epsilon\epsilon\,\delta^{d-n+1}\Pi^2\Pi^{n-2}
=(d-n+1)\epsilon\epsilon\,\delta^{d-n}\Pi^n .
\ee
as desired.

\bibliographystyle{JHEP}
\bibliography{references}

\end{document}